\documentclass[sigconf]{acmart}
\renewcommand\footnotetextcopyrightpermission[1]{}
\usepackage{cuted}
\usepackage{paralist}
\usepackage{todonotes}
\usepackage{enumerate}
\usepackage{verbatim}
\usepackage[table]{xcolor}
\usepackage{xspace}
\usepackage{caption}
\usepackage{hyperref}
\usepackage{cleveref}
\usepackage{booktabs}
\usepackage{tabularx}
\usepackage{makecell}
\usepackage{multirow}
\usepackage{colortbl}
\usepackage{pifont}
\usepackage{pdflscape}
\usepackage{threeparttable}
\usepackage{array}
\usepackage{cleveref}

\Crefname{lstlisting}{listing}{listings}
\Crefname{lstlisting}{Listing}{Listings}
\usepackage{listings}
\usepackage{tcolorbox}

\definecolor{LLM_Web_Agent_C}{HTML}{FEA4E6}
\definecolor{LLM_Web_Agent_L}{HTML}{FF8C66}
\definecolor{Automation_Frameworks}{HTML}{2ACCFE}
\definecolor{CLI_Scrapers}{HTML}{58BF03}
\definecolor{HumanColor}{HTML}{999999}
\definecolor{RowGray}{gray}{0.96}

\definecolor{lowentropy}{HTML}{A3B7FF}   
\definecolor{highentropy}{HTML}{EBC80D}  

\definecolor{zerogray}{gray}{0.75}
\newcommand{\zero}{\textcolor{zerogray}{0}}

\definecolor{rulecolor}{RGB}{52,152,219}      
\definecolor{captchacolor}{RGB}{46,204,113}   
\definecolor{powcolor}{RGB}{230,126,34}       
\definecolor{cloudcolor}{RGB}{155,89,182}     

\definecolor{kvkey}{HTML}{DD1144}
\definecolor{kvvalue}{HTML}{0c6e82}

\lstdefinelanguage{kv}{
    basicstyle=\ttfamily\small,
    columns=fullflexible,
    keepspaces=true,
    moredelim=[s][\color{kvkey}]{'}{=},
    moredelim=[s][\color{kvvalue}]{"\$}{"}
}

\lstdefinelanguage{prompt}{
  morekeywords={Task, Input, Output},
  sensitive=false
}

\newtcolorbox{takeawaysbox}{
  colback=gray!10,
  colframe=black,
  title=Key Takeaways,
  fonttitle=\bfseries,
  boxrule=0.5pt,
  arc=2pt,
  left=6pt,
  right=6pt,
  top=6pt,
  bottom=6pt
}

\newcommand{\cmark}{\textcolor{green!50!black}{\ding{51}}}
\newcommand{\blocked}{\textcolor{red!70!black}{\ding{55}}} 
\newcommand{\timeout}{\textcolor{orange!90!black}{\ding{108}}} 

\newcommand{\CH}{\texttt{Client Hello}\xspace}
\newcommand{\LLMb}{LLM-based\xspace}

\newcommand{\cur}{\emph{cURL}\xspace}
\newcommand{\wge}{\emph{wget}\xspace}
\newcommand{\scr}{\emph{scrapy}\xspace}
\newcommand{\Sel}{\emph{Selenium}\xspace}
\newcommand{\Play}{\emph{Playwright}\xspace}
\newcommand{\Pup}{\emph{Puppeteer}\xspace}
\newcommand{\OC}{\emph{OpenClaw}\xspace}
\newcommand{\Claude}{\emph{Claude}\xspace}
\newcommand{\CAI}{\emph{Crawl4AI}\xspace}
\newcommand{\CAIs}{\emph{Crawl4AI-Stealth}\xspace}
\newcommand{\CAIub}{\emph{Crawl4AI-Undetected-Browser}\xspace}
\newcommand{\BU}{\emph{BrowserUse}\xspace}
\newcommand{\BUs}{\emph{BrowserUse-Stealth}\xspace}
\newcommand{\Chat}{\emph{ChatGPT Agent}\xspace}
\newcommand{\Sky}{\emph{Skyvern}\xspace}

\newcommand{\Rtxt}{\texttt{robots.txt}\xspace}
\newcommand{\WA}{Web Agent\xspace}
\newcommand{\WAs}{Web Agents\xspace}
\newcommand{\antibot}{anti-bot mechanism\xspace}
\newcommand{\antibots}{anti-bot mechanisms\xspace}

\newcommand{\etal}{et~al.\xspace}
\newcommand{\ie}{i.e.,\xspace}
\newcommand{\eg}{e.g.,\xspace}

\begin{document}

\title[On the Internet, Nobody Knows You're an LLM Bot]{On the Internet, Nobody Knows You're an LLM Bot:\\
Unmasking Web Agents with Multi-Layer Fingerprinting}

\begingroup
\renewcommand{\thefootnote}{\fnsymbol{footnote}}

\author{Iliana Fayolle}
\authornote{These authors contributed equally to this work.}
\orcid{0009-0000-6690-529X}
\affiliation{%
  \institution{Univ. Lille, CNRS, Inria}
  \city{Lille}
  \country{France}}
\email{iliana.fayolle@inria.fr}

\author{Sihem Bouhenniche}
\authornotemark[1]
\orcid{0009-0000-0175-9092}
\affiliation{%
  \institution{Univ. Lille, CNRS, Inria}
  \city{Lille}
  \country{France}}
\email{sihem.bouhenniche@inria.fr}

\endgroup
\setcounter{footnote}{0}

\author{Samuel Pélissier}
\orcid{0000-0002-3554-2585}
\affiliation{%
  \institution{CentraleSupélec, Inria, IRISA}
  \city{Rennes}
  \country{France}}
\email{samuel.pelissier@inria.fr}

\author{Pierre Laperdrix}
\orcid{0000-0001-6901-3596}
\affiliation{%
  \institution{Univ. Lille, CNRS, Inria}
  \city{Lille}
  \country{France}}
\email{pierre.laperdrix@inria.fr}

\author{Clémentine Maurice}
\orcid{0000-0002-8896-9494}
\affiliation{%
  \institution{Univ. Lille, CNRS, Inria}
  \city{Lille}
  \country{France}}
\email{clementine.maurice@inria.fr}

\author{Walter Rudametkin}
\orcid{0000-0003-2903-7600}
\affiliation{%
  \institution{Univ. Rennes, CNRS, Inria, IRISA, IUF}
  \city{Rennes}
  \country{France}}
\email{walter.rudametkin@inria.fr}

\renewcommand{\shortauthors}{Fayolle et al.}

\begin{abstract}
Since 2023, a new class of bots has emerged: \WAs.
They can automate complex tasks on the Web, going beyond traditional browser automation tools such as \emph{Selenium}, \emph{Puppeteer}, or \emph{Playwright}. Leveraging \emph{large language models} (LLMs), these agents are capable of solving \antibots, mimicking human behavior, and, in some cases, operating directly from the local machine of the user configuring them.
As a result, it is becoming increasingly difficult for website administrators to detect and block these LLM-based bots. 
Modern \WAs commonly integrate stealth and anti-detection techniques, while numerous proprietary and open-source \antibots have emerged recently, specifically to block them.
However, despite their growing prevalence, there is little evaluation of the effectiveness of state-of-the-art \antibots against these \LLMb bots and their stealth capabilities. Likewise, no prior work has comprehensively studied how to characterize and distinguish \WAs deployed either in the cloud or locally.

This paper addresses these open questions by deploying multiple honeysites protected by one or more \antibots (\eg \Rtxt, CAPTCHAs, proof-of-work, and Cloudflare's free proprietary solutions). We integrated network-, HTTP-, and browser-level fingerprinting techniques, and prompted six \LLMb \WAs to visit the deployed honeysites.
Our analysis reveals three main findings:
\textbf{(i)} some \WAs were able to bypass all evaluated \antibots;
\textbf{(ii)} all evaluated \WAs can be distinguished both from humans and from one another using multi-layer fingerprinting techniques across network, HTTP and browser layers;
\textbf{(iii)} stealth and anti-detection mechanisms often increase detectability rather than decrease it.
\end{abstract}


\maketitle

\section{Introduction}

According to Cloudflare Radar~\cite{CloudflareRadarbots} and the 2025 Thales Group report~\cite{Bad_bots_rise_internet_traffic_hits_record_levels}, bots represent between 30\% and 50\% of all web traffic.
Not all bots are harmful, some serve legitimate purposes in the Web ecosystem, such as search engine indexers, uptime monitors, or accessibility checkers.
However, a significant fraction of bot traffic is either \emph{malicious} (\eg engaging in security exploits, credential stuffing, fraud, spam, or content theft~\cite{chiapponi2020hopla, Ad_Account_Farming_Agent, DBLP:journals/corr/abs-2411-15091}) or simply \emph{unwanted} (\eg consuming server resources, inflating bandwidth costs, and crowding out genuine user traffic~\cite{Please_stop_externalizing_your_costs_directly_into_my_face}). 
Approximately 14\% of total web traffic is attributed to so-called \emph{bad bots}~\cite{Bad_bots_rise_internet_traffic_hits_record_levels}. 
As a result, web administrators need to selectively block certain categories of bots while preserving access for legitimate bots and actual visitors.

This dynamic explains the ongoing arms race between bot operators and web administrators.
The most basic countermeasure, the Robots Exclusion Protocol (commonly known as \Rtxt~\cite{Introduction_to_robots.txt}), allows web administrators to declare which paths crawlers should avoid.
However, since \Rtxt is merely advisory, many bots, including some of those operated by AI companies, ignore it entirely and scrape content regardless~\cite{DBLP:journals/corr/abs-2411-15091, Table_of_bot_metrics}. 
Moreover, non-expert web administrators, such as artists, hobbyists, or beginners are often unaware of this convention or are unable to modify the file due to restrictions imposed by their hosting provider~\cite{DBLP:journals/corr/abs-2411-15091}.
More direct enforcement mechanisms, such as server-side rules~\cite{Apache_HTTP_Server} and user-agent blocking, shift control from the bot to the server but remain trivially bypassed via user-agent spoofing~\cite{Web_Bot_Auth, DBLP:conf/sp/LiARN21, Perplexity_AI_Is_Lying}. 
Hence, more sophisticated techniques have been developed, including challenge-based methods (\eg CAPTCHAs~\cite{reCAPTCHA, Pro-Captcha}, proof-of-work~\cite{Anubis}), browser~\cite{vastel:hal-02441653, DBLP:journals/corr/abs-2406-07647, DBLP:conf/dimva/AzadSLN20} and TLS fingerprinting~\cite{DBLP:conf/iscc/PapadogiannakiI23, DBLP:conf/sp/LiARN21}, which aim to identify bots more reliably regardless of the identity they claim. 

Since October 2023, a new category of bot traffic has emerged, based on Large Language Models (LLMs), a class of AI generative models~\cite{DBLP:journals/corr/abs-2411-15091}.
\LLMb bots operate in three distinct modes: \textbf{(i)}~\emph{passive crawling} to collect training data from the open Web~\cite{Overview_of_OpenAI_Crawlers}; \textbf{(ii)}~\emph{active retrieval} triggered when a user submits a query to a system such as ChatGPT~\cite{ChatGPT}, causing the model to fetch live content on the user's behalf; and \textbf{(iii)}~\emph{agentic visits} in which autonomous \WA tools browse, interact with, and extract information from pages with varying degrees of autonomy (hereafter, we use the terms ``tools'' and \WAs interchangeably).
This increase in \LLMb traffic has concrete consequences for website administrators: excessive resource consumption, increased hosting fees, loss of advertising revenue from traffic never converted into human engagement, and a loss of control over how content is used for model training~\cite{DBLP:journals/corr/abs-2411-15091}.
Beyond volume, \LLMb bots require more nuance when blocking.
A web administrator could previously maintain a relatively stable allowlist of well-known, well-behaved crawlers (\eg \emph{Googlebot}) and broadly block any unexpected behavior or fingerprint.
Now, they may wish to block LLM training crawlers to prevent their content from being ingested without consent, while simultaneously permitting certain agentic use cases.

However, previous work only focused on \Rtxt compliance by \LLMb bots~\cite{DBLP:conf/imc/KimBLLPW25, DBLP:journals/corr/abs-2411-15091, DBLP:conf/ccs/Cui00L25} or the characteristics of bot traffic prior to LLM democratization~\cite{DBLP:conf/sp/LiARN21, DBLP:journals/corr/abs-2406-07647, vastel:hal-02441653, vastel2018fp, DBLP:conf/dimva/AzadSLN20}.
To the best of our knowledge, this is the first large-scale study on the identifying features of \LLMb bots from the perspective of a web server using only single-request, multi-layer fingerprinting techniques via technical API-level observations, without relying on behavioral analysis that requires multiple bot visits or interactions~\cite{wang2026fp}.
Furthermore, we found no previous work that estimates the possibility of blocking evasive \LLMb bots with a sufficiently fine-grained approach, which is crucial for web administrators to maintain control over their content and resources while allowing legitimate use cases.
This motivates the following research questions:

\textbf{RQ1}: To what extent are modern web defenses effective against bots, in particular \LLMb ones?

\textbf{RQ2}: Do \LLMb bots exhibit distinctive characteristics (\eg IP metadata, TLS fingerprints, HTTP header patterns, browser fingerprints) that differentiate them from traditional bots?

\textbf{RQ3}: To what extent can locally executed \WAs be distinguished from cloud-based ones, and are they inherently less detectable due to their similarity to genuine user environments?

To answer these research questions, we deployed nine honeypot servers, each with one or more \antibots installed. For each incoming connection, we collect three fingerprinting layers: \textbf{(i)} network-level information including IP and TLS data, \textbf{(ii)} HTTP headers, and \textbf{(iii)} browser fingerprints. We interact with our infrastructure through a diverse set of 12 tools, including historical web scrapers (\eg \cur), web automation frameworks (\eg \Play), and \LLMb \WAs, deployed both locally and in cloud environments. To establish a baseline for legitimate traffic, we additionally collect fingerprints generated by our local machines across the same three layers.

Our contributions are as follows. \textbf{(I)} We actively evaluate the effectiveness of each \antibot, as well as the added value of combining multiple techniques against modern \WAs (§~\ref{sec:anti_bot_defenses_evaluation}).
\textbf{(II)} We establish the most relevant fingerprinting criteria to distinguish between human users and bots, and more specifically to differentiate between \WAs and traditional bots (§~\ref{sec:characterizing_bots}).
\textbf{(III)} Finally, we create an anonymized four-month dataset containing passive and active fingerprinting data from various categories of bots.
Using the active data, we evaluate multi-layer classification and achieve accurate classification of bots, \LLMb bots, and human users (§~\ref{sec:multi-layer_classification}).

\noindent\emph{Artifacts are available in the \nameref{sec:OS} section.}

\section{Background}
In this section, we present the different types of bots, \antibots, and fingerprinting techniques necessary to understand the contributions of this paper.

\subsection{Taxonomy of Bots}
\label{sec:background:typesofbots}

Bots automate a variety of tasks on the Web, including crawling, scraping, and interacting with applications. We classify them into three categories based on their level of automation and underlying technology stack: lightweight HTTP scrapers, browser automation frameworks, and \LLMb \WAs.

\paragraph{\textbf{HTTP-based scrapers}}
The most basic bots rely on HTTP clients and scraping tools such as \cur~\cite{curl}, \wge~\cite{wget}, or \scr~\cite{scrapy}.
Because they operate outside a browser environment, these bots cannot execute JavaScript or render dynamic web applications; they are limited to retrieving raw HTML. 
Nevertheless, this approach is widely adopted for large-scale data collection due to its minimal computational overhead and reduced hosting costs.
In practice, training crawlers like \emph{GPTBot}~\cite{Overview_of_OpenAI_Crawlers} aggressively scan websites to harvest data, sometimes revisiting dynamic pages multiple times a day. This pattern can inadvertently lead to outages cause DDoS-like disruptions~\cite{Please_stop_externalizing_your_costs_directly_into_my_face}.
Similarly, when users prompt an \LLMb chatbot to retrieve website information, the assistant may dispatch an HTTP-based scraper, such as \emph{ChatGPT-User}~\cite{Overview_of_OpenAI_Crawlers}, to fetch the content in real time~\cite{DBLP:conf/ccs/Cui00L25}.

\paragraph{\textbf{Browser automation frameworks}} Frameworks such as \Sel~\cite{selenium}, \Pup~\cite{puppeteer}, or \Play~\cite{playwright} allow bots to control genuine web browsers to visit websites.
Unlike HTTP-based tools, they execute JavaScript and render dynamic webpages, making them harder to detect because they are closer to standard user environments.
They mimic human browsing through scripted interactions (\eg clicking, filling forms, scrolling), but often struggle with non-deterministic challenges like CAPTCHAs.

\paragraph{\textbf{\WAs}}
\LLMb bots operate at the semantic layer, following natural language instructions instead of requiring fully scripted interactions~\cite{OpenSourceWebAgents, ChatGPT_Agent, Crawl4AI, OpenClaw, Anthropic_Claude_For_Chrome, BrowserUse, Skyvern}.
Most \WAs combine browser automation frameworks with \LLMb reasoning.
Early systems focused on text-based browsing~\cite{nakano2021webgpt}, whereas recent agents can visually interpret webpages~\cite{he2024webvoyager, Skyvern, ChatGPT_Agent}, adapt to JavaScript-heavy environments~\cite{Skyvern, ChatGPT_Agent, Anthropic_Claude_For_Chrome}, and even control desktop environments~\cite{OpenClaw}.
The growing sophistication of \WAs raises major concerns for website administrators.
They can extract high-value content at scale~\cite{DBLP:journals/corr/abs-2411-15091}, automate abusive workflows, such as credential stuffing or ticket scalping~\cite{Meta_Takes_Down_Inauthentic_Accounts_on_Facebook, Ad_Account_Farming_Agent, chiapponi2020hopla}, and increasingly integrate anti-detection mechanisms, including stealth browser configurations~\cite{BrowserUse_Stealth, Crawl4AI_Undetected_Browser}. 
As a result, modern \WAs are substantially harder to detect than traditional bots because \textbf{(i)} they execute full browser stacks, \textbf{(ii)} they exhibit realistic interaction patterns~\cite{DBLP:journals/corr/abs-2506-01952, DBLP:journals/tmlr/YuZLGYWZYNHC26, BrowserUse_Cloud}, \textbf{(iii)} they rely on vision-based reasoning~\cite{he2024webvoyager, Skyvern, ChatGPT_Agent}, and \textbf{(iv)} they may operate locally on the users device~\cite{OpenClaw}.

\subsection{Anti-bot Mechanisms}
\label{sec:botprot}
To mitigate bots, websites rely on defense strategies that vary in popularity, openness, and effectiveness.
A common approach is to rely on IP address reputation by blocking addresses previously associated with suspicious behavior.
However, this alone is insufficient, as demonstrated in this paper due to bots easily rotating IP addresses (see §~\ref{sec:ip_layer}), and must therefore be combined with other \antibots.

\paragraph{\textbf{Rule-based detection}}
Basic defenses include the \Rtxt convention~\cite{Introduction_to_robots.txt, What_is_robots.txt?} and server-side configuration files, such as Apache's \verb|.htaccess|~\cite{Apache_HTTP_Server}.
Formally known as the Robots Exclusion Protocol~\cite{rfc9309, DBLP:conf/imc/KimBLLPW25}, \Rtxt specifies which resources crawlers should avoid through directives such as \texttt{Disallow} and \texttt{User-agent}.
However, compliance is voluntary~\cite{DDoS_from_Anthropic_AI}, especially for malicious bots, and empirical studies show strong variations across crawlers~\cite{DBLP:conf/imc/KimBLLPW25}.
While many \LLMb crawlers follow these directives, \WAs often do not~\cite{DBLP:journals/corr/abs-2411-15091, DBLP:conf/imc/KimBLLPW25}.
More specific conventions such as the \verb|noai| meta tag also exist but are rarely adopted~\cite{DBLP:journals/corr/abs-2411-15091}.

Server-side configurations (\eg \verb|.htaccess| rules, firewalls) provide a stronger mechanism by blocking specific user agents or IP addresses, typically returning errors like the HTTP 403 status code.
Respectful \LLMb crawlers, including those from OpenAI~\cite{Overview_of_OpenAI_Crawlers}, often identify themselves through their user agent and can therefore be filtered~\cite{DBLP:journals/corr/abs-2411-15091}.
Nevertheless, bots can evade these rules by spoofing user agents or rotating IP addresses~\cite{DBLP:conf/sp/LiARN21}.

\paragraph{\textbf{CAPTCHAs}}
Completely Automated Public Turing tests to tell Computers and Humans Apart (CAPTCHAs) aim to distinguish humans from bots.
Over the past two decades, they have evolved from text recognition to images, puzzles, videos, audio, equations, and mini-games~\cite{DBLP:journals/csur/GuerarVMPM22}.
However, advances in machine learning have enabled bots to solve many traditional CAPTCHAs, especially simpler ones~\cite{How_CAPTCHAs_work, DBLP:journals/corr/abs-2405-07496, DBLP:journals/corr/abs-2006-08296, 9742729}. Additionally, attackers bypass them using CAPTCHA farms, where underpaid workers manually solve the challenges for bots~\cite{CAPTCHA_Farms}.
As a result, more advanced approaches have emerged, including checkbox-based ("I'm not a robot") and frictionless CAPTCHAs, which rely on fingerprinting and behavioral analysis instead of explicit user interaction.
A widely deployed example is Google's \emph{reCAPTCHA}~\cite{reCAPTCHA}, which analyzes signals such as cursor movements, cookies, device history, and browsing behavior to detect automation. Privacy-friendly and GDPR-compliant alternatives also exist, such as \emph{ProCaptcha}~\cite{Pro-Captcha}.

\paragraph{\textbf{Proof-of-work}}
Proof-of-work mechanisms require a client to solve a computational challenge before accessing a website~\cite{DBLP:conf/cms/JakobssonJ99}.
This increases the cost of large-scale crawling by forcing bots to use sufficiently powerful hardware.
Unlike CAPTCHAs, proof-of-work does not require user interaction or personal data.
For example, \emph{Anubis}~\cite{Anubis} asks clients to compute SHA-256 checksums before granting access.
Although lightweight, open-source, and free, such systems have limitations: sophisticated bots may execute the challenge in full browser environments or outsource it to external servers capable of solving SHA-256 puzzles~\cite{Anubis_bypassed}.
As acknowledged in the official Anubis documentation~\cite{Anubis}, proof-of-work should therefore complement stronger approaches like fingerprinting (see~§~\ref{sec:fingerprinting}).

\paragraph{\textbf{Cloudflare solutions}}
\emph{Cloudflare}~\cite{Cloudflare} provides widely used hosting and security services for websites. 
According to W3Techs~\cite{Usage_statistics_and_market_share_of_Cloudflare}, 22.7\% of websites used Cloudflare as of May 2026, making it a major actor in bot mitigation.
Their \antibots include: \textbf{(i)}~\emph{Turnstile}~\cite{Cloudflare_Turnstile}, a privacy-focused alternative to CAPTCHAs that silently analyzes browser behavior, JavaScript execution, and device characteristics through fingerprinting to estimate whether a visitor is human~\cite{sateur2025evaluating}.
\textbf{(ii)}~\emph{Bot Fight Mode (BFM)}~\cite{Cloudflare_Bot_Fight_Mode}, a free proof-of-work-based system that detects suspicious traffic and serves computationally expensive JavaScript challenges.
Paid versions provide additional controls but rely on the same detection technology~\cite{Cloudflare_Super_Bot_Fight_Mode}.
\textbf{(iii)}~\emph{Block AI Bots}~\cite{Declare_your_AIndependence}, which blocks known \LLMb crawlers used for AI training, including those from \emph{ChatGPT}~\cite{ChatGPT}, \emph{Claude}~\cite{Claude}, \emph{Perplexity}~\cite{Perplexity}, and \emph{Gemini}~\cite{Gemini}.
Cloudflare maintains lists of associated User-Agent strings and IP addresses, including bots that do not identify themselves~\cite{Declare_your_AIndependence}.
Although the implementation details remain private, Liu \etal~\cite{DBLP:journals/corr/abs-2411-15091} empirically analyzed the AI crawlers blocked by this feature and observed that \LLMb crawlers operated by major AI companies generally respected \Rtxt directives, whereas several \LLMb applications did not consistently do so.

\subsection{Fingerprinting}
\label{sec:fingerprinting}
Anti-bot mechanisms often rely on \emph{fingerprinting} to distinguish bots from legitimate users.
In practice, fingerprinting consists of collecting discriminative attributes and comparing them to known bot signatures or behaviors~\cite{husak2016HTTPSTrafficAnalysis, laperdrix2020browser}.
This section presents fingerprinting approaches at both the network and browser levels.

\paragraph{\textbf{Transport Layer Security (TLS) fingerprinting}}

Transport Layer Security~\cite{rfc8446} is a communication protocol to secure network communications, such as HTTPS.
At the start of a TLS connection, the client sends an unencrypted \CH message advertising supported cipher suites, TLS versions, extensions, and cryptographic preferences.
Because the structure and ordering of these elements depend on the client software and TLS library, the \CH message is often characteristic of a specific software stack~\cite{husak2016HTTPSTrafficAnalysis, jarad2026handshakes}.
A widely adopted fingerprinting tool is JA4~\cite{ja4_blog, ja4_cloudflare_blog, ja4_GTI_blog, jarad2026handshakes}, which extracts structured features from the \CH message to generate a concise fingerprint (\ie a hash) that can be used across datasets.
Since our work focuses on identifying bot characteristics rather than improving fingerprinting itself, we use JA4 as an off-the-shelf solution.

\paragraph{\textbf{Browser fingerprinting}}

Although browser fingerprinting is widely known for device tracking~\cite{laperdrix2020browser,segura2018operation, liu2025first}, it is also used for bot detection~\cite{vastel:hal-02441653}.
In practice, JavaScript code embedded in webpages collects distinctive browser and device attributes, such as software versions, screen resolution, fonts, plugins, and HTTP headers~\cite{laperdrix2020browser}, and combines them into a quasi-unique identifier~\cite{eckersley2010unique}.
Because fingerprinting is stateless and does not rely on client-side storage, it can persist across sessions and resist evasion techniques like user-agent spoofing~\cite{DBLP:conf/sp/LiARN21, DBLP:journals/corr/abs-2406-07647, vastel2018fp, jarad2026handshakes}.
It is also effective at detecting inconsistencies or anomalous configurations associated with automation~\cite{DBLP:journals/corr/abs-2406-07647, vastel2018fp}, such as unusual font sets, missing plugins, or mismatched system attributes.

\section{Methodology} \label{sec:methodo}

Our goal is to evaluate the effectiveness of \antibots against \WAs while simultaneously assessing the ability of \WAs to evade these defenses by mimicking human characteristics. 
In particular, we analyze: 
\textbf{(i)} the differences between various categories of bots, such as traditional scrapers and autonomous \WAs, and 
\textbf{(ii)} the differences across multiple layers of collected information, including IP metadata, TLS fingerprints, HTTP headers, and browser fingerprints.
To achieve this, we designed and deployed an infrastructure based on \emph{honeysites} to collect real-world data both passively (\ie from web scrapers), and actively (\ie by instrumenting \WAs).
In this section, we describe our \emph{honeysites} infrastructure, our experimental protocol, as well as the data processing and metrics we adopted.

\subsection{Infrastructure}

\begin{figure*}[h!]
    \centering
    \includegraphics[width=0.9\linewidth]{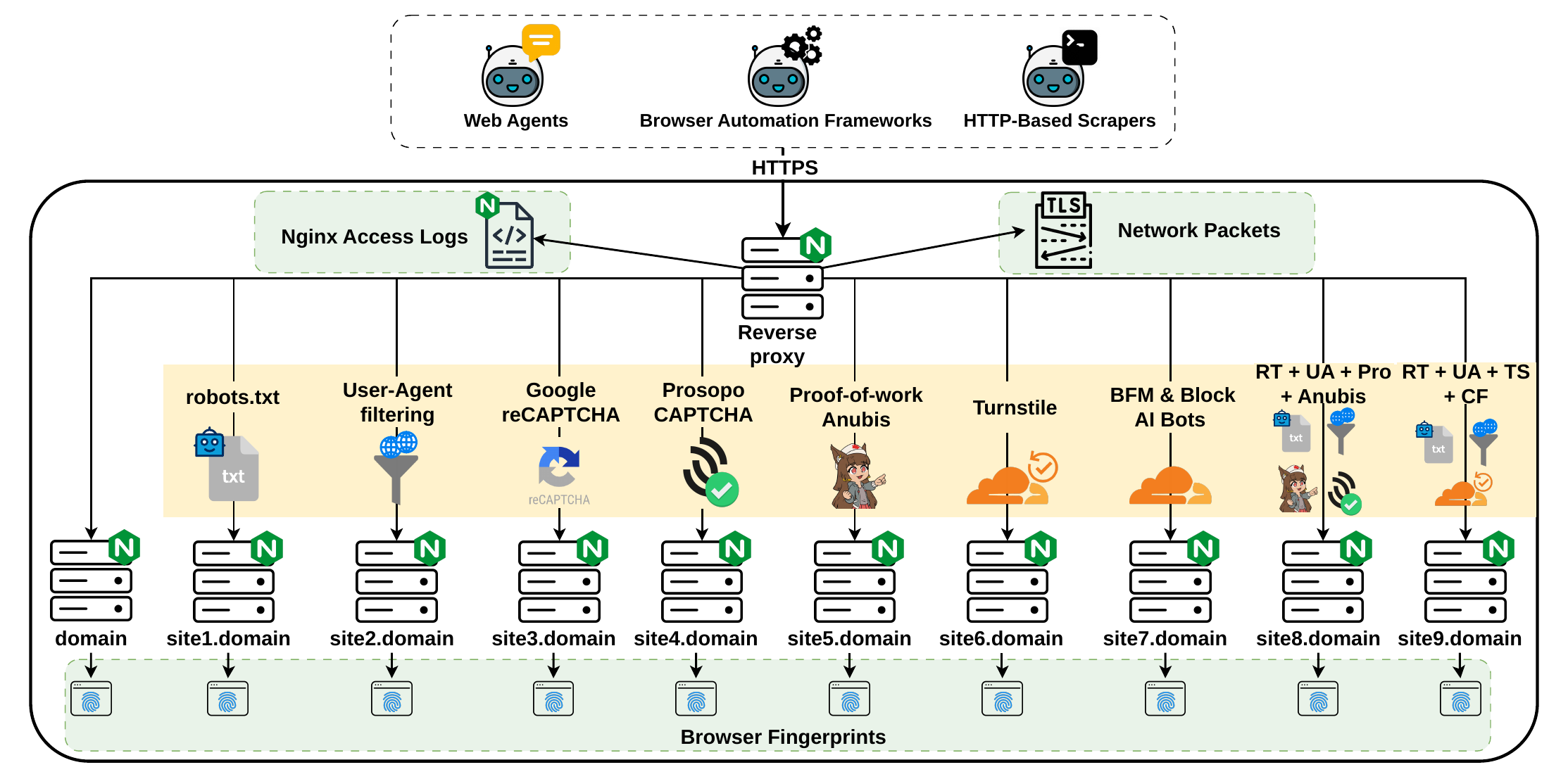}
    \caption{Data collection infrastructure for the deployment and instrumentation of honeysites.}
    \Description{Data collection infrastructure overview, where traffic generated by web agents, browser automation frameworks, and HTTP-based scrapers passes through a reverse proxy that records Network Packets and Nginx Access Logs before forwarding requests to multiple subdomains protected by different anti-bot mechanisms. Browser Fingerprints are collected from each subdomain website.}
    \label{fig:infrastructure}
\end{figure*}

As summarized in \Cref{fig:infrastructure}, we rely on \emph{honeysites} with different \antibots that capture signals across network and application layers, allowing us to study bot interactions with web environments. 

\subsubsection{Anti-bot Mechanisms Across Honeysites.}
\label{sub:honey_antibot}
The defenses we selected (summarized in \Cref{tab:honeysites_protection}) are widely used and representative of the state-of-the-art, covering all categories of anti-bot mechanisms introduced in §~\ref{sec:botprot}.
For \Rtxt and \emph{User-Agent filtering}, we used the bot lists recommended in~\cite{rtxt_block, rtxt_UA_block}, reflecting common blocking practices. Additional details are provided in the artifacts in the \nameref{sec:OS} section.
We excluded \LLMb \antibots, such as \emph{AI Labyrinth}~\cite{AI_Labyrinth}, as their objective is not to block bots, but to redirect visitors identified as bots by Cloudflare's \antibots toward AI-generated decoy pages designed to waste their time and computational resources.
We also investigate whether combining \antibots increases bot detection.
To this end, we implemented two combined configurations: one relying exclusively on free and open-source solutions, and another based on Cloudflare protections.
In both cases, we retained rule-based mechanisms because they are lightweight, easy to deploy, and free to use.

\begin{table}[t]
\caption{Selected \antibots.}
\centering

\begin{threeparttable}

\renewcommand{\arraystretch}{1.05}

\begin{tabular}{p{5.5cm} c}
\toprule
\textbf{Defense(s)} & \textbf{Availability} \\
\midrule

\colorbox{rulecolor!15}{\Rtxt~(RT)} &
 Open standard \\

\colorbox{rulecolor!15}{User-Agent filtering~(UA)} &
 Open-source \\

\colorbox{captchacolor!15}{Google reCAPTCHA v3 (rV3)*} &
Proprietary \\

\colorbox{captchacolor!15}{Prosopo CAPTCHA v3.5 \& v3.6~(Pro)} &
Open-source \\

\colorbox{powcolor!15}{Proof-of-work Anubis v1.20.0-pre1~(Anubis)} &
Open-source \\

\colorbox{cloudcolor!15}{Turnstile~(TS)*} &
Proprietary \\

\colorbox{cloudcolor!15}{
BFM* \& Block AI Bots~(CF)*} &
Proprietary \\

\colorbox{rulecolor!15}{RT} +
\colorbox{rulecolor!15}{UA} +
\colorbox{captchacolor!15}{Pro} +
\colorbox{powcolor!15}{Anubis} &
Open-source \\

\colorbox{rulecolor!15}{RT} +
\colorbox{rulecolor!15}{UA} +
\colorbox{cloudcolor!15}{TS} +
\colorbox{cloudcolor!15}{CF} &
Proprietary  \\

\bottomrule
\end{tabular}

\begin{tablenotes}[flushleft]
\footnotesize
\item
\colorbox{rulecolor!15}{Rule-based detection}
\colorbox{captchacolor!15}{CAPTCHA}
\colorbox{powcolor!15}{Proof-of-work}
\colorbox{cloudcolor!15}{Cloudflare solutions}
\item \textbf{*} indicates that the version of the corresponding solutions is not publicly specified. Experiments were conducted January--May 2026.
\end{tablenotes}
\end{threeparttable}
\label{tab:honeysites_protection}
\end{table}

\subsubsection{Honeysite Deployment}
\label{sec:honeysite_deployment}

Each \textit{honeysite} is deployed on its own nginx web server with zero, one, or multiple \antibots listed in \Cref{tab:honeysites_protection}. 
An nginx reverse proxy routes incoming traffic to the appropriate domains.
All honeysites are served over HTTPS and support TCP-based protocols (HTTP/1.1 and HTTP/2) as well as UDP-based HTTP/3 (QUIC).
To ensure each visit generates a fresh request, HTTP caching is disabled.
To prevent visits from unsuspecting human users, the honeysites are hosted under a randomized domain.
The main domain hosts the unprotected honeysite, while additional honeysites are deployed behind the \texttt{site1}--\texttt{site9} subdomains.

\subsubsection{Honeysite Design}

We designed our honeysites using a combination of static and dynamic elements to capture a broad range of bot behaviors. Figure \ref{fig:clockblog} in Appendix~\ref{appendix:website_appearance} illustrates the appearance of the \emph{honeysites}, which feature a large clock that updates regularly along with relevant textual information. The static content is intended to create an authentic environment for indexing bots and simple scrapers that do not execute JavaScript. The dynamic content, including shuffled menu links, changing clock times, and a dynamically generated unique page identifier (\texttt{page\_id}),
is designed to verify that \WAs interact with a live page rather than a cached version that could bias our measurements.
To reliably identify \WAs, we require them to submit a three-field HTML form (first name, last name, and content) via an HTTP POST request. Each interaction is tracked using a session cookie (\texttt{cookie\_id}) assigned during the initial visit. 
We consider an agent’s visit successful if it bypasses our security measures and correctly executes the submission by providing: the form data including the dynamic \texttt{page\_id} in the POST payload, and the \texttt{cookie\_id} via the HTTP cookie header.

\subsubsection{Collected Data}
\label{sub:collected_data}
For each visit to a \emph{honeysite}, we collect data types from three connection layers:

\noindent \textbf{(i)}~\emph{Network packets}.
We capture IP and TLS packets of all incoming traffic at the entry point of our infrastructure using the network protocol analyzer \texttt{tshark}~\cite{Tshark}. This allows us to collect traffic from both the TCP and UDP protocols.
\textbf{(ii)}~\emph{Nginx access logs}.
We retain nginx access logs~\cite{LogType} at the entry point for a complete view of the visits.
\textbf{(iii)}~\emph{Browser fingerprints}.
We extract browser fingerprints using a lightweight algorithm inspired by AmIUnique~\cite{amiunique} and FingerprintJS~\cite{FingerprintJS}. 
The fingerprint contains JavaScript attributes and HTTP request headers, listed in Appendix~\ref{appendix:Attributes}.

\subsection{Experimental Protocol}

We generate visits to our \emph{honeysites} using different \WAs and browser automation frameworks. Though not the focus of this paper, we also monitored passive traffic (see Appendix~\ref{appendix:passive_data}).

\paragraph{\textbf{Local machines}}
For all local operations, we used two Intel Core i7-1185G7 machines (running Ubuntu 20.04.6 and 24.04.3) to contrast regular browsing behavior with automation tools.

\paragraph{\textbf{Human users}}
To establish a ground-truth baseline, we manually accessed the \emph{honeysites} from both local machines using Chrome (v144.0.7559.96) and Firefox (v136.0 and v147.0.1).

\paragraph{\textbf{Scrapers and browser automation frameworks}}

We generated traffic from our local machines using scrapers (\emph{cURL}, \emph{wget}, \emph{scrapy}) and browser automation frameworks (\emph{Selenium}, \emph{Playwright}, \emph{Puppeteer}), as detailed in §~\ref{sec:background:typesofbots}.

\paragraph{\textbf{\WAs}}

\begin{table*}
\centering
\small
\rowcolors{2}{gray!10}{white}
\caption{Feature comparison of selected \WAs.}
\begin{tabular}{lccccccc}
\toprule
\textbf{Agent} 
& \makecell{Proprietary\\vs open source} 
& \makecell{Local\\vs cloud} 
& \makecell{Anti-bot\\capabilities} 
& Browser 
& \makecell{Automation\\Framework} 
& \makecell{Browser\\version} 
& \makecell{Agent\\version} \\
\midrule

OpenClaw ~\cite{OpenClaw} 
& Open source 
& Local 
& \makecell{Undocumented\\(stealth mention in source code)}
& Chrome 
& Playwright 
& 144.0.7559.96 
& 2026.2.2-3 \\

Claude for Chrome ~\cite{Anthropic_Claude_For_Chrome} 
& Proprietary
& Local 
& Undocumented 
& Chrome 
& Extension 
& 144.0.7559.96 
& 1.0.40 \\

Crawl4AI ~\cite{Crawl4AI} 
& Open source
& Local 
& \makecell[c]{Stealth mode \\ Undetected Browser mode} 
& \makecell[c]{Chromium } 
& Playwright 
& 145.0.7632.6
& 0.8.0 \\

BrowserUse ~\cite{BrowserUse} 
& Mixed 
& Both 
& Stealth mode 
& \makecell[c]{Chrome \\ Chromium} 
& Playwright 
& \textdagger
& 0.11.\{5, 6, 9\} \\

ChatGPT Agent ~\cite{ChatGPT_Agent} 
& Proprietary
& Cloud 
& Undocumented 
& \makecell[c]{Chrome \\ Chromium} 
& Undocumented 
& 141.0.7390.122 
& * \\

Skyvern ~\cite{Skyvern} 
& Proprietary
& Cloud
& Stealth mode 
& Edge 
& Playwright 
& \makecell{[143.0.3650.139 \\ 144.0.3719.92]}
& 1.0.10 \\

\bottomrule
\end{tabular}
\begin{tablenotes}
\footnotesize
\begin{minipage}{\textwidth}
\item \textdagger~The list of minor versions of \BU is:
Google Chrome 144.0.7559.\{0, 59, 60, 61, 96, 97, 98, 109, 110\} and Chromium 145.0.7632.6.
\item *~We did not find the version number, so we instead report the experiment dates: 01-28-2026, 02-08-2026, and 02-09-2026.
\end{minipage}
\end{tablenotes}
\label{tab:webagents}
\end{table*}

As previous work~\cite{OpenSourceWebAgents, ukani2025privacy}, we selected six \LLMb \WAs based on their popularity (\ie GitHub stars).
They are either cloud-based, deployed locally (as web extensions or via Playwright), or both.
The \WA selection is summarized in \Cref{tab:webagents} and more details on their features are given in Appendix~\ref{appendix:WA_details}.
We excluded general-purpose AI assistants like \emph{ChatGPT}~\cite{ChatGPT}, \emph{Gemini}~\cite{Gemini}, and \emph{Perplexity}~\cite{Perplexity}, since our experiments showed they typically retrieve content via simplified HTTP requests, similar to \cur. As a result, they do not execute client-side scripts or perform user-like actions, making them trivial to detect or block.

We instructed the \WAs to visit our \emph{honeysites} several times, which let us collect multiple fingerprints per \WA and analyze their stability.
We followed official documentation and used the default configurations. 
Although commercial \WAs do not publicly document their anti-detection techniques, we identified stealth mechanisms in all open-source solutions.
Moreover, \CAI and \BU advertise bot-detection bypassing capabilities in their official documentation, further motivating the need to characterize \LLMb bot features.
Consequently, when anti-detection mechanisms are available, we repeated the experiment with the anti-detection mode enabled.
Each visit is driven by predefined, structured prompts provided in Appendix~\ref{appendix:prompts}, inspired by prior work on evaluating \LLMb browsing behavior~\cite{DBLP:journals/corr/abs-2411-15091, DBLP:conf/ccs/Cui00L25}. 
\emph{Normal prompts} instruct the agent to visit the targeted \emph{honeysite}, interact with the webpage, and return a structured response containing the task status and the dynamic \texttt{page\_id}. 
In addition, we use a \emph{special prompt} variant instructing the agent to modify its behavior when blocked to evaluate whether the agent is capable of bypassing \antibots.

For cloud-based \WAs, we manually monitored the tasks as they did not expose an API suitable for our automation pipeline. For locally deployed agents, we used their APIs to automate the experiments on our machines.
Since \CAI is designed primarily for web crawling rather than task-oriented interactions, we monitored experiments using a simplified version of the \textit{normal prompt}, asking it only to fetch the \texttt{page\_id} displayed on the page.

\subsection{Data Processing and Metrics}

After data collection, we map individual visits with all collected data (\ie network packets, nginx access logs, and browser fingerprints). 
Then, we analyze each layer using dedicated metrics to assess the attributes discriminative power and answer \textbf{RQ2} and \textbf{RQ3}. 

\subsubsection{Fingerprint Layer Association}

Because our data is collected asynchronously across multiple layers using disparate tools, we developed a multi-step linking process to correlate visits with their respective fingerprints.
First, we link each visit to its browser fingerprint using the \texttt{page\_id}.
Second, we associate browser fingerprints with nginx access logs using the \texttt{IP address}, \texttt{User-Agent} header, and \texttt{cookie\_id}
within a two-minute window of launching the task and fingerprint collection. 
Finally, we associate network packets with their corresponding nginx access logs using \texttt{<source IP address, source port>} tuples within a five-minute interval corresponding to the default SSL session timeout~\cite{SSL_get_default_timeout}.
Ultimately, this process yields a unified profile for each visit that encompasses its network, and browser fingerprints.

\subsubsection{Metrics}
\label{subsub:attribute_analysis}

To identify discriminative characteristics, we analyze each layer and identify attributes that remain consistent within a tool category (\ie \WAs, scrapers, automation frameworks, humans) while differing across categories.
We first clean and normalize the dataset into a unified representation where each column corresponds to a single attribute.
Attributes constant across all visits are removed since they provide no discriminative value.
We then evaluate attributes using probability-based metrics instead of entropy-based commonly used in fingerprinting, since it mainly quantifies attribute variability rather than capturing attribute consistency and distinctiveness within a given category.
Furthermore, entropy-based metrics are biased on small datasets, as attribute values may appear highly distinctive and stable for a given category while still exhibiting low entropy due to their small size.
This limitation is mitigated by the metrics described below.

\textbf{Intra-tool Value Probability} is the probability that the attribute $A$ takes the value $v$ in fingerprints of the tool $t$: \[ Intra(A, v, t) = P(A = v \mid t) \]

\noindent When $Intra(A,v,t)$ is close to 1, the value $v$ is consistently observed for the attribute $A$, whereas low values indicate that $v$ is rare or inconsistent.

\textbf{Inter-tool Value Exclusivity} measures how distinctive a value $v$ of attribute $A$ is for tool $t$, by considering its occurrences in fingerprints generated by other tools ($t' \neq t$):
\[ Inter(A, v, t) = 1 - P(A = v \mid t' \neq t) \]

\noindent When $Inter(A,v,t)$ is close to 1, the value $v$ is rare across other tools and therefore highly specific to tool $t$.
Conversely, lower values indicate that $v$ is commonly observed in other tools.

\textbf{V-score} combines the two previous metrics and quantifies how strongly a value $v$ of attribute $A$ characterizes tool $t$: 
\[ V\text{-}Score(A, v, t) = Intra(A, v, t) \times Inter(A, v, t) \]

\textbf{A-Score} measures the overall discriminative power of an attribute $A$ by aggregating the contributions of all its observed values: 
\[ A\text{-}Score(A, t) = \sum_{v \in Vals(A, t)} V\text{-}Score(A, v, t) \]

\noindent A high $A\text{-}Score(A,t)$ indicates that attribute $A$ consistently exhibits one or more values that are both stable and exclusive for tool $t$.

\section{Dataset}

We generated traffic to our \emph{honeysites} between January 28 and February 24, 2026, and collected the associated fingerprinting data. 
Out of the 1,449 initiated active visits, further detailed in Appendix~\ref{appendix:active_data}, some had to be excluded from the final dataset due to \antibots or technical errors. 
Cloudflare blocked visits before reaching our servers for \BU (40), HTTP-based scrapers (20), and \Pup (1).
In addition, \OC generated three timeouts while trying to access the \emph{honeysites}. 
After excluding failed visits, we successfully associated 1,385 visits with their corresponding nginx access logs and browser fingerprints.
When associating visits with TLS packets, two visits could not be matched with any TLS packets as their IPs were not found in our captures and were thus excluded.
Finally, our active visits dataset contains 1,383 visits associated with 1,383 browser fingerprints and 1,358 TLS \CH records.
The difference arises because multiple visits may share the same TLS session, and thus the same \CH.

\section{Evaluating \antibots}
\label{sec:anti_bot_defenses_evaluation}

\begin{table*}[t]
\centering
\footnotesize
\caption{Bot access to honeysites under different bot defense mechanisms.}
\label{tab:agents_against_defenses}
\begin{threeparttable}
\setlength{\tabcolsep}{2.8pt}
\renewcommand{\arraystretch}{0.95}
\rowcolors{2}{RowGray}{white}
\begin{tabularx}{\textwidth}{l l l *{10}{>{\centering\arraybackslash}X}}
\toprule
\textbf{Tool} & \textbf{Model} & \textbf{Infra.} &
\textbf{ND} & \textbf{RT} & \textbf{UA} & \textbf{rV3} & \textbf{Pro} & \textbf{Anubis} & \textbf{TS} & \textbf{CF} & \textbf{A} & \textbf{B} \\
\midrule

\rowcolor{HumanColor!25}
Humans & -- & -- &
\cmark & \cmark & \cmark & \cmark & \cmark & \cmark & \cmark & \cmark & \cmark & \cmark \\

\rowcolor{CLI_Scrapers!15}
cURL & -- & L &
\cmark & \cmark & \cmark & \blocked & \blocked & \blocked & \blocked & \cmark & \blocked & \blocked \\

\rowcolor{CLI_Scrapers!20}
wget & -- & L &
\cmark & \cmark & \cmark & \blocked & \blocked & \blocked & \blocked & \cmark & \blocked & \blocked \\
\rowcolor{CLI_Scrapers!25}
scrapy & -- & L &
\cmark & \blocked & \blocked & \blocked & \blocked & \blocked & \blocked & \cmark & \blocked & \blocked \\

\rowcolor{Automation_Frameworks!15}
Selenium & -- & L &
\cmark & \cmark & \cmark & \cmark & \blocked & \cmark & \blocked & \cmark & \blocked & \blocked \\

\rowcolor{Automation_Frameworks!20}
Playwright & -- & L &
\cmark & \cmark & \cmark & \cmark & \blocked & \cmark & \blocked & \cmark & \blocked & \blocked \\

\rowcolor{Automation_Frameworks!25}
Puppeteer & -- & L &
\cmark & \cmark & \cmark & \cmark & \blocked & \cmark & \blocked & \cmark & \blocked & \blocked \\

\rowcolor{LLM_Web_Agent_L!15}
OpenClaw & sonnet-4.5 & L &
\cmark & \cmark & \cmark & \cmark & \cmark & \cmark & \cmark & \cmark & \cmark & \cmark\ \\

\rowcolor{LLM_Web_Agent_L!15}
OpenClaw & opus-4.5 & L &
\cmark & \cmark & \cmark & \cmark & \cmark & \cmark & \cmark & \cmark & \cmark & \cmark \\

\rowcolor{LLM_Web_Agent_L!20}
Claude Chrome & sonnet-4.5 & L &
\cmark & \cmark & \cmark & \cmark & \cmark & \cmark & \cmark & \cmark & \cmark & \cmark \\

\rowcolor{LLM_Web_Agent_L!20}
Claude Chrome & opus-4.5 & L &
\cmark & \cmark & \cmark & \cmark & \blocked & \cmark & \cmark & \cmark & \blocked & \cmark \\

\rowcolor{LLM_Web_Agent_L!30}
Crawl4AI & gpt-4o-mini & L &
\cmark & \cmark & \cmark & \cmark & \blocked & \cmark & \cmark & \blocked & \blocked & \blocked \\

\rowcolor{LLM_Web_Agent_L!30}
Crawl4AI Stealth & gpt-4o-mini & L &
\cmark & \cmark & \cmark & \blocked & \blocked & \cmark & \blocked & \cmark & \blocked & \blocked \\

\rowcolor{LLM_Web_Agent_L!30}
Crawl4AI Undetected & gpt-4o-mini & L &
\cmark & \cmark & \cmark & \blocked & \blocked & \cmark & \cmark & \cmark & \blocked & \cmark \\

\rowcolor{LLM_Web_Agent_L!40}
BrowserUse & bu-1-0 & L &
\cmark & \cmark & \cmark & \cmark & \blocked & \cmark & \cmark & \cmark & \blocked & \cmark \\

\rowcolor{LLM_Web_Agent_L!40}
BrowserUse & sonnet-4.5 & L &
\cmark & \cmark & \cmark & \cmark & \blocked & \cmark & \cmark & \cmark & \blocked & \cmark \\

\rowcolor{LLM_Web_Agent_C!15}
BrowserUse & bu-2-0 & C &
\cmark & \cmark & \cmark & \cmark& \blocked & \cmark & \cmark & \cmark & \blocked & \blocked \\

\rowcolor{LLM_Web_Agent_C!15}
BrowserUse & sonnet-4.5 & C &
\cmark & \cmark & \cmark & \cmark & \blocked & \cmark & \cmark & \cmark & \blocked & \cmark \\

\rowcolor{LLM_Web_Agent_C!23}
BrowserUse Stealth & bu-1-0 & C &
\cmark & \cmark & \cmark & \blocked & \blocked & \cmark & \cmark & \blocked & \blocked & \blocked \\

\rowcolor{LLM_Web_Agent_C!23}
BrowserUse Stealth & sonnet-4.5 & C &
\cmark & \cmark & \cmark & \blocked & \blocked & \cmark & \cmark & \blocked & \blocked & \blocked \\

\rowcolor{LLM_Web_Agent_C!30}
ChatGPT Agent & CUA & C &
\cmark & \cmark & \cmark & \cmark& \cmark & \cmark & \blocked & \cmark & \cmark & \blocked \\

\rowcolor{LLM_Web_Agent_C!40}
Skyvern & Opti. & C &
\cmark & \cmark & \cmark & \blocked & \timeout & \cmark & \cmark & \cmark & \blocked & \cmark \\

\rowcolor{LLM_Web_Agent_C!40}
Skyvern & gpt-5.2 & C &
\cmark & \cmark & \cmark & \blocked & \timeout & \cmark & \cmark & \cmark & \blocked & \cmark \\

\bottomrule
\end{tabularx}
\begin{tablenotes}
\scriptsize
\item Colors matching:
\colorbox{CLI_Scrapers!40}{HTTP-based scrapers} \quad
\colorbox{Automation_Frameworks!40}{Browser automation frameworks} \quad
\colorbox{LLM_Web_Agent_L!40}{\WAs local} \quad
\colorbox{LLM_Web_Agent_C!40}{\WAs cloud}
\item \textbf{ND}: No Defense \quad \textbf{RT}: \Rtxt \quad \textbf{UA}: User-Agent filtering \quad \textbf{rV3}: reCAPTCHA v3 \quad \textbf{Pro}: Prosopo CAPTCHA v3.5 \& v3.6 \quad \textbf{Anubis}: Anubis v1.20.0-pre1 \quad \textbf{TS}: Turnstile \quad \textbf{CF}: BFM \& Block AI Bots
\item \textbf{A}: RT + UA + Pro + Anubis \quad \textbf{B}: RT + UA + TS + CF \quad \cmark Success \quad \blocked Blocked \quad \timeout Timeout \quad Infra.: Infrastructure \quad L: Local Browser \quad C: Cloud Browser 
\item Results were obtained using either a normal prompt or, when the normal prompt failed, a special prompt.
\end{tablenotes}
\end{threeparttable}
\end{table*}

To answer \textbf{RQ1}, we evaluate the success of various tools in bypassing different \antibots described in §~\ref{sub:honey_antibot}.
\Cref{tab:agents_against_defenses} summarizes the results of our evaluation across multiple tools, models, and \antibots.
We also assess whether combining multiple defenses (\textbf{A} and \textbf{B}) is more effective than deploying them individually, due to their multi-layered approach.

Overall, the effectiveness of \antibots varies significantly across tools and models.
\OC and \Claude for Chrome, using the \texttt{Sonnet 4.5} model, \emph{reliably bypass all protections}.
In contrast, using a less capable model, \Claude for Chrome with \texttt{Opus 4.5}, successfully identified the correct interaction area of the Prosopo CAPTCHA but failed to execute the required click action. 
Combining multiple defense mechanisms appears effective in blocking the majority of the automated tools.
However, accumulating multiple CAPTCHA challenges may significantly degrade usability for legitimate users.
The use of special prompts (see Appendix~\ref{appendix:prompts}) did not improve the overall success rate of the tests.

\paragraph{\textbf{HTTP-based scrapers}}
Only \scr was blocked by the \Rtxt and User-Agent filtering rules.
This behavior results from \emph{scrapy}’s default configuration through the \verb|ROBOTSTXT_OBEY = True| setting, which enforces \Rtxt compliance.
Since our file explicitly restricted access, \scr respected these rules and stopped crawling.
The other HTTP-based scrapers did not request the \Rtxt file before crawling, and were neither explicitly restricted User-Agent filtering rules.
Hence, such defenses rely primarily on scraper goodwill and are therefore unlikely to be effective against malicious actors in real-world scenarios.

As expected, HTTP-based bots were successfully blocked by Anubis' proof-of-work, which requires JavaScript to solve hash-based challenges. 
Additionally, we observed that Anubis inspects the User-Agent string prior to issuing the challenge, which likely contributed to the blocking since the scrapers used default, identifiable User-Agent values.
On the contrary, Cloudflare BFM did not block HTTP-based bots.
After carefully reviewing the logs, we found that it is correctly installed and running, but only blocks traffic coming from known datacenters.
As our tests were conducted from university IP addresses, we believe their reputation was high enough to not trigger \antibots despite clearly advertising automation-related User-Agents.

\paragraph{\textbf{Browser automation frameworks}}
Automation frameworks successfully passed threshold-based reCAPTCHA v3, due to using configurations similar enough to those of genuine browsers.
However, automation scripts must account for each \antibot in their various versions and will fail on unexpected behaviors.
For instance, when Prosopo CAPTCHA or Cloudflare Turnstile flagged the visits and presented a challenge, browser automation frameworks were unequipped to go through.

\paragraph{\textbf{\WAs}} 
The main observations for \WAs are twofold: 
\textbf{(i)} a small number of agents successfully bypass all evaluated defenses, while 
\textbf{(ii)} agents advertising stealth or anti-detection capabilities do not necessarily achieve better results and may even become easier to detect.

\OC and \Claude Chrome using the \texttt{sonnet-4.5} model are the only evaluated tools that successfully bypass all tested defenses.
In particular, both tools successfully solved the Prosopo CAPTCHA, which proved effective against most other agents because solving the challenge requires not only reasoning about the task, but also reliable interacting with the verification mechanism.
Indeed, these tools and \Chat were able to solve the challenge by directly interacting with the verification button without requiring additional scripting or repeated prompting.
However, \Chat was still blocked by Turnstile, likely due to specific HTTP header characteristics that prevented rapid verification and left the page indefinitely in a verification state.
Although Turnstile did not explicitly deny access, the prolonged verification delay caused the agent either to abandon the navigation attempt or to exhibit alternative bypass behaviors, as described in Appendix \ref{sec:notable_behavior}.

Several agents also provide stealth or anti-detection configurations, but these mechanisms did not consistently improve bypass capabilities and sometimes increased detectability instead.
In contrast to \OC and \Claude Chrome, stealth-enabled variants of \CAI and \BU remained detectable by multiple protections.
Enabling stealth mode made both tools more suspicious to frictionless CAPTCHA systems.
For \emph{Crawl4AI}, we observed substantial variations in HTTP headers that were absent during non-stealth executions.
Since the stealth implementation is derived from \emph{Playwright}, a framework already heavily targeted by \antibots, this increased detection rate is not entirely surprising.
In stealth mode, additional browser-configuration overrides, such as \verb|magic|, \verb|simulate_user| and \verb|override_navigator|, modify native browser properties and may unintentionally increase fingerprint inconsistencies, including against Turnstile.
For \BUs mode, on its own cloud infrastructure to function properly in this mode, resulting in traffic originating from blacklisted addresses by \emph{Cloudflare} and \emph{Google}.
We give more details on attributes and values that may flag bots in the following section.

\section{Characterizing Bots}
\label{sec:characterizing_bots}

To answer \textbf{RQ2} and \textbf{RQ3}, we analyze the fingerprinting characteristics of different bots interacting with our honeysite.
Our analysis focuses on three main layers of the connection stack, ordered by how quickly they can be accessed: the IP layer, the TLS fingerprinting layer, and finally the browser fingerprinting layer.
We then evaluate the effectiveness of combining these layers for bot detection and characterization. 

\subsection{IP Layer}
\label{sec:ip_layer}
\begin{table}[t]
\centering
\footnotesize
\setlength{\tabcolsep}{4pt}
\renewcommand{\arraystretch}{1.1}
\caption{IP-layer characteristics per tool.}
\begin{tabular}{l|lcc}
\toprule
\textbf{Tool} & \textbf{\#ASN} & \textbf{IP Hosting} & \textbf{IP Abuser} \\
\midrule

\multicolumn{4}{l}{\textbf{Humans and Local Tools}} \\
\cellcolor{HumanColor!20} Human & 1 & 0.00 & 0.00 \\
\cellcolor{CLI_Scrapers!20} HTTP-based Scrapers\textbf{*} & 1 & 0.00 || 0.25 & 0.00 || 0.25 \\ 
\cellcolor{Automation_Frameworks!20} Browser Automation Framework & 1 & 0.00 & 0.00 \\ 
\cellcolor{LLM_Web_Agent_L!20} BrowserUse & 1 & 0.00 & 0.00 \\
\cellcolor{LLM_Web_Agent_L!25} Other local \WAs & 1 & 0.00 & 0.00 \\
\specialrule{\lightrulewidth}{1pt}{1pt}

\multicolumn{4}{l}{\textbf{Cloud \WAs}} \\
\cellcolor{LLM_Web_Agent_C!20} BrowserUse & 35 & 0.02 & 0.02 \\
\cellcolor{LLM_Web_Agent_C!25} BrowserUse-Stealth & 3 & 0.98 & 0.98 \\
\cellcolor{LLM_Web_Agent_C!30} ChatGPT Agent  & 1 & 0.88 & 0.14 \\
\cellcolor{LLM_Web_Agent_C!35} Skyvern & 34 &  0.01 & 0.00 \\

\bottomrule
\end{tabular}

\begin{tablenotes}
\scriptsize
\item \textbf{\#ASN}: number of distinct Autonomous Systems observed for a given tool.
\item \textbf{IP Hosting}: proportion of IPs identified as hosting/datacenter infrastructure.
\item \textbf{IP Abuser}: proportion of IPs flagged by the \emph{IPLocate}~\cite{IPLOCATE} IP reputation service as abusive \\or suspicious.
\item \textbf{*} Variations are mainly caused by Cloudflare \antibots acting as a proxy. \\
For tools without a browser context, the observed IP address corresponds to Cloudflare\\ infrastructure because these tools cannot persist the \texttt{cf\_clearance} cookie~\cite{cf-Clearance}\\ required to pass Cloudflare challenges, causing the server to observe traffic generated\\
by Cloudflare instead of traffic from the tools. 
\end{tablenotes}
\label{tab:ip_layers}
\end{table}

The IP layer provides the first signals for bot detection and characterization.
Although IP reputation labels and hosting classifications may vary across providers and over time, the relative differences between tools remained consistent throughout our measurements.
We leverage the \emph{IPLocate}~\cite{IPLOCATE} IP reputation service for labeling our dataset.
Results for this layer are shown in \Cref{tab:ip_layers}.

While results presented in \Cref{tab:ip_layers} follow our expectations, they also illustrate specific deployment decisions and anti-bot evasion practices.
First, local-based tools leverage the user's IP address, taking advantage from its already good reputation and relevant location.
Second, cloud-based \WAs can opt for two routing options.
They directly connect from datacenters, \eg \Chat always visited our honeysites from the \emph{Cloudflare} infrastructure located in the European Union.
These IP addresses are already flagged and may increase bot detection rates.
Alternatively, we observe connections coming from numerous ASNs with high quality IP addresses, which corresponds to US-based residential proxies advertised by \BU~\footnote{More surprisingly, \BUs mode avoids their residential proxy infrastructure and directly connects from low-reputation datacenter IP addresses. It is unclear wether this is due to a deployment misconfiguration or intended, as it increases bot detection.} and \emph{Skyvern}.

Although well-established datacenter IP addresses remain a significant signal, this behavior effectively reduces the relevance of IP addresses alone~\cite{mehanna2024FreeProxiesUnmaskeda} and further motivates our analyzes of upper network layers.   

\subsection{TLS Layer}
\label{sec:tls_layer}

\begin{table}[t]
\centering
\footnotesize
\setlength{\tabcolsep}{4pt}
\renewcommand{\arraystretch}{1.1}
\caption{Dominant JA4 and discriminative power per tool.}
\begin{tabular}{l|lccc}
\toprule
\textbf{Tool (\& Browser Version)} & \textbf{JA4 Index} & \textbf{Intra.} & \textbf{V.} & \textbf{A.} \\
\midrule

\multicolumn{4}{l}{\textbf{Humans (Local)}} \\
\cellcolor{HumanColor!20} Firefox 136.0         & 5\textdagger, \textbf{6}\textdagger, 16 & 0.60 & 0.60 & 0.99 \\
\cellcolor{HumanColor!20} Firefox 147.0.1       & \textbf{7}, 8 & 0.50 & 0.50 & 0.99 \\
\cellcolor{HumanColor!20} Chrome 144.0.7559.96  & 1, \textbf{3}, 11 & 0.45 & 0.38 & 0.82 \\
\specialrule{\lightrulewidth}{1pt}{1pt}

\multicolumn{4}{l}{\textbf{HTTP-based Scrapers (Local)}} \\
\cellcolor{CLI_Scrapers!20} cURL 8.15.0-DEV  & 9, \textbf{20} & 0.75 & 0.75 & 1.00 \\
\cellcolor{CLI_Scrapers!20} wget 1.21.4    & 9, \textbf{21}\textdagger & 0.75 & 0.75 & 1.00 \\
\cellcolor{CLI_Scrapers!20} scrapy 2.14.1    & \textbf{19}\textdagger & 1.00 & 1.00 & 1.00 \\
\specialrule{\lightrulewidth}{1pt}{1pt}

\multicolumn{4}{l}{\textbf{Browser Automation (Local)}} \\
\cellcolor{Automation_Frameworks!20} Selenium (Firefox 147.0.2)       & 7, 16, \textbf{17} & 0.78 & 0.78 & 0.99 \\
\cellcolor{Automation_Frameworks!20} Selenium (Chrome 144.0.7559.96)  & \textbf{1}, 3, 11 & 0.48 & 0.40 & 0.83 \\

\cellcolor{Automation_Frameworks!25} Playwright (Firefox 146.0)       & 7, 8, \textbf{16} & 0.72 & 0.70 & 0.97 \\
\cellcolor{Automation_Frameworks!25} Playwright (Chromium 145.0.7632.6) & \textbf{1}, 3, 11 & 0.48 & 0.40 & 0.83 \\

\cellcolor{Automation_Frameworks!30} Puppeteer (Firefox 147.0.3)        & 7, 8, \textbf{16}, 17 & 0.71 & 0.69 & 0.97 \\
\cellcolor{Automation_Frameworks!30} Puppeteer (Chromium 145.0.7632.46) & \textbf{2}\textdagger, 4\textdagger, 14\textdagger, 15\textdagger & 0.46 & 0.46 & 1.00 \\

\specialrule{\lightrulewidth}{1pt}{1pt}

\multicolumn{4}{l}{\textbf{\WAs (Local)}} \\
\cellcolor{LLM_Web_Agent_L!20} OpenClaw (Chrome 144.0.7559.96)   & 1, \textbf{3} & 0.59 & 0.52 & 0.86 \\
\cellcolor{LLM_Web_Agent_L!20} OpenClaw (python-request)           & \textbf{20} & 1.00 & 1.00 & 1.00 \\

\cellcolor{LLM_Web_Agent_L!25} Claude (Chrome 144.0.7559.96)     & 1, \textbf{3}, 11 & 0.60 & 0.53 & 0.86 \\

\cellcolor{LLM_Web_Agent_L!30} Crawl4AI (Chromium 145.0.7632.6)        & 1, 3, \textbf{11} & 0.76 & 0.63 & 0.82 \\
\cellcolor{LLM_Web_Agent_L!30} Crawl4AI-Stealth                             & 1, \textbf{11} & 0.80 & 0.66 & 0.83 \\
\cellcolor{LLM_Web_Agent_L!30} Crawl4AI-Undetected-Browser                           & 1, \textbf{11} & 0.78 & 0.64 & 0.82 \\

\cellcolor{LLM_Web_Agent_L!35} BrowserUse (Chrome 144) & \textbf{1}, 3, 11 & 0.56 & 0.49 & 0.85 \\
\specialrule{\lightrulewidth}{1pt}{1pt}

\multicolumn{4}{l}{\textbf{\WAs (Cloud)}} \\
\cellcolor{LLM_Web_Agent_C!20} BrowserUse &\textbf{10}\textdagger, 12 & 0.98 & 0.89 & 0.91 \\
\cellcolor{LLM_Web_Agent_C!25} BrowserUse-Stealth &\textbf{10}\textdagger, 12 & 0.98 & 0.88 & 0.90 \\

\cellcolor{LLM_Web_Agent_C!30} ChatGPT (Chromium 141) & \textbf{18}\textdagger & 1.00 & 1.00 & 1.00 \\

\cellcolor{LLM_Web_Agent_C!35} Skyvern (Edge 143 \& 144) & 11, \textbf{13}\textdagger & 0.52 & 0.45 & 0.93 \\

\bottomrule
\end{tabular}

\begin{tablenotes}
\scriptsize
\item \textbf{Bold}: dominant JA4 fingerprint. \quad \textdagger: unique to the tool.
\item Intra.: proportion of fingerprint with the dominant JA4.
\item V.: discriminative dominant JA4 V-Score (Intra-Score $\times$ Inter-Score).
\item A.: discriminative A-score (sum of V-Score for all JA4 of the tool).
\item Mapping between JA4 index and JA4 hash and its distribution are provided in Appendix~\ref{appendix:JA4}.
\end{tablenotes}
\label{tab:ja4_ascore}
\end{table}

The TLS \CH packet is one of the earliest fingerprintable elements visible after the IP layer during HTTPS exchanges.
According to Salesforce’s original JA3/JA4 publication~\cite{JA3_JA3S_TLS_fingerprinting_overview}, TLS fingerprints such as JA4 are generally more stable than IP-based identifiers because they reflect the client TLS implementation rather than network location.
Compared to browser-based fingerprinting, this enables lightweight early-stage bot detection before higher-layer protocol interactions occur.
In this study, we leverage JA4 fingerprints~\cite{ja4, ja4_blog} to characterize the TLS behavior of each tool and evaluate the stability and distinctiveness of these fingerprints across configurations and repeated executions.

In practice, we find 21 unique JA4 fingerprints. For brevity, \Cref{tab:ja4_ascore} summarizes the corresponding indexes of each observed JA4 values, the dominant JA4 value in bold (\ie the most frequently observed fingerprint), its uniqueness (marked with \textdagger), its \emph{Intra-tool Value Probability} or \emph{Intra-Score} (\ie the proportion of connections exhibiting this dominant JA4 for the considered tool), its \emph{V-Score}, and the \emph{A-Score}.
A V-Score close to 1 indicates that a tool is associated with a highly distinctive dominant JA4, whereas a score close to 0 reflects low distinctiveness.
An A-Score close to 1 indicates that the JA4 of a tool are both stable and unique across observations, while a score close to 0 suggests that the tool cannot be reliably distinguished based on its fingerprints alone. 
The complete mapping between the 21 JA4 indexes, their respective JA4 hashes, and detailed distributions are provided in Appendix~\ref{appendix:JA4}.

\paragraph{\textbf{Multiple factors influence JA4 fingerprint variability}}

Observed JA4 values are primarily influenced by the underlying client’s TLS implementation, such as NSS for Firefox~\cite{NSS} or BoringSSL for Chromium-based browsers~\cite{BoringSSL}.
For instance, the JA4 fingerprints observed between Firefox~136 and Firefox~147.0.1 differ only by the presence of the \texttt{compress\_certificate} extension~\cite{rfc8879}, likely due to NSS version updates~\cite{NSS3108, NSS3119}.
It may also stem from browser-level configuration or user preferences~\cite{Editor_Mozilla, Editor_Chrome}.
Moreover, a single tool may exhibit multiple JA4 values depending on the transport protocol or TLS session state. 
For example, when possible, some tools may use QUIC instead of TCP to accelerate connection establishment, resulting in distinct JA4 fingerprints. 
Finally, not all observed TLS extensions may be standardized yet or publicly documented.
Some experimental extensions contribution to the JA4 fingerprint may be absent from IANA registries, RFC specifications, and thus \texttt{tshark} dissectors used for this study.

We observe a few specific behaviors in our dataset. For instance, \Chat only establishes TCP connections and advertises 28 cipher suites, which is more than most evaluated tools, while exposing only 12 TLS extensions, fewer than most others.
This behavior suggests a design choice favoring broad compatibility with websites relying on older or heterogeneous TLS configurations.
A similar strategy can be observed for HTTP-based scrapers, which generally rely exclusively on TCP connections, advertise many cipher suites, and expose relatively few extensions, likely to maximize compatibility across a wide range of target websites.
On the opposite, \textit{Crawl4AI} exhibits two dominant JA4 values: one associated with QUIC connections and another with TCP connections.
Similarly, the presence of the \texttt{pre\_shared\_key} (PSK) extension~\cite{rfc8773}, used for TLS session resumption, may generate different JA4 values depending on whether a session is resumed or newly established. 
This behavior is observed for \textit{Skyvern}, which presents two dominant JA4 fingerprints: one with the PSK extension and one without.

\paragraph{\textbf{Cloud-based tools and HTTP-based scrapers are more distinguishable from humans than local-based tools}}
According to the $Intra\text{-}$ and $A\text{-}Scores$, distinguishing individual cloud-based agents from one another is generally easier than distinguishing local-based agents from each other or from human-operated browsers.
Most cloud-based agents also exhibit a dominant JA4 value unique to the corresponding tool, whereas local-based agents frequently share their dominant JA4 values with human browser configurations and generally achieve lower Intra-Scores.
This behavior is expected, as cloud-based agents typically operate in tightly controlled and homogeneous environments, leading to more stable TLS fingerprints.
In contrast, local-based agents inherit variability from the host browser and operating system configuration used by the human operator.

\paragraph{\textbf{Stealth or undetected modes do not significantly reduce detectability}}

Neither the use of stealth prompts nor the choice of underlying model had a significant impact on the observed JA4 fingerprints.
Configurations using stealth or undetected modes exhibited dominant JA4 values and A-Scores similar to those observed without such configurations, indicating that these mechanisms do not substantially alter the TLS fingerprinting surface.

\subsection{Browser Fingerprinting Layer}
\label{sub:bf_layer}
Browser features represent the last technical layer used for fingerprinting bots. Some values are available directly in HTTP headers (\eg User Agent), but most are extracted after page load using JavaScript (\eg screen resolution), and thus require a full-fledge browser.
Our dataset contains 1,383 browser fingerprints, among which 13 do not include JavaScript-extracted attributes because it was not enabled.
Overall, automated tools exhibit high $A\text{-}Scores$ for specific attributes, indicating stable and distinctive configurations. We provide further details in Table~\ref{tab:top_discriminative_bf_attributes} of Appendix~\ref{appendix:attributes_bf} for completeness. 
In the following, we discuss broad groups of attributes which are both consistent and discriminative across tools. 
 
\paragraph{\textbf{User Agent}}
While local tools such as \Claude and \OC expose the underlying local browser versions, cloud-based \WAs exhibit distinctive patterns. For example, \Sky is the only tool using Microsoft Edge, whereas \Chat exposes a unique Chrome version and platform combination. In contrast, \BU cloud and stealth modes reveal multiple \emph{User Agent} values, reflecting the diversity of their cloud environments. As one of the most discriminative attributes in our dataset, User Agent is also easily modified by bots. However, we can compare possibly altered HTTP header values with the more stable \texttt{userAgentData} JavaScript API~\cite{userAgentData} to detect spoofing, which is an additional identifying signal. Using this method, we found that \CAI attempts to spoof its HTTP \emph{User Agent}, replacing its actual version with a default, older value~\cite{BrowserCrawlerLLM}. 

\paragraph{\textbf{Screen and display}} 
These attributes, including width and height, differ significantly across tools.  
While some \WAs like \textit{Crawl4AI} or \textit{BrowserUse} allow manual configuration, other cloud-based tools only offer immutable environments with identifiable default values. For example, \Chat is configured with a fixed screen resolution of $1280 \times 960$, having a $V\text{-}Score$ of $1$, which means that it is the only tool exhibiting this resolution in our dataset.

\paragraph{\textbf{Permissions state}} We observed notable discrepancies in permission states (\eg access to sensors or notifications) across tools. 
This attribute achieves high $A\text{-}Scores$ ($>0.70$) for all studied tools, indicating strong discriminative power.
For instance, the majority of permission queries through the API \texttt{navigator.permissions-}\\\texttt{.query} fail for \CAI and \Chat, which may indicate that many browser capabilities are unavailable.
In contrast, \BU consistently returns a \texttt{denied} state for most permission queries, indicating a more restrictive behavior by default compared to \Sky and \Claude, which more closely aligns with what we observe for our human-based visits.
These differences reflect how each tool initializes and configures the browser context, and often remain stable across executions.

\paragraph{\textbf{Cookie management}} Handling cookies is influenced by the tools' designs. 
For instance, extensions like \Claude can inherit cookies from the user’s browser. Likewise, \Chat, \Sky, and \OC share cookies across multiple sessions to use a continuous user identity. 
On the contrary, \CAI and \BU use fresh browsing sessions, unless explicitly configured.

\paragraph{\textbf{Number of CPU cores}} 
Although local tools report a stable configuration of 8 CPU cores, corresponding to the authors' architecture, we obtained high $A\text{-}Score$ values for cloud-based \WAs. 
Indeed, these consistently expose distinctive values. For instance, \Sky reports 32 CPU cores and \Chat 13 CPU cores, while \BU cloud infrastructure exhibits high variability, with the number of reported CPU cores ranging from 2 to 64. While it only reflects the number of logical cores exposed by virtualized environments, it remains a strong signal.

\paragraph{\textbf{Stealth behavior}} In line with the degraded evasion capacity presented in §~\ref{sec:anti_bot_defenses_evaluation}, we found that "stealth" configurations can introduce detectable inconsistencies in browser fingerprints.
For example, \CAI stealth mode modifies the User Agent and injects synthetic \texttt{referer} values pointing to external websites. 
These changes create atypical attribute combinations rarely observed in genuine human browsing, making them discriminative signals. 

\paragraph{\textbf{Tool-specific attributes}} Some tools expose specific signals that uniquely characterize them. 
For instance, \CAI stealth mode reports the string \texttt{"default"} for the \texttt{notifications} permission, which is not a standard permission state. 
More legitimately yet easily fingerprintable, \textit{ChatGPT Agent} advertises its identity using two HTTP headers, \texttt{signature} and \texttt{signature-agent}~\cite{backman2024HTTPMessageSignatures}.

Overall, \textit{ChatGPT Agent} and \textit{Skyvern} are the easiest tools to distinguish. 
They rely on fixed execution environments, resulting in highly stable attributes that are not observed in other tools.
Moreover, standard automation frameworks including \textit{Selenium}, \textit{Playwright}, and \textit{Puppeteer} remain consistently detectable, primarily due to well-known signals, \eg \texttt{navigator.webdriver}. 
While this alone is sufficient to identify automated behavior, additional attributes such as screen dimensions further differentiate these tools from each other, reflecting differences in their default configurations.
In contrast, tools implementing stealth mechanisms such as \textit{Crawl4AI} and \textit{BrowserUse} exhibit more diverse fingerprints, but this introduces numerous inconsistencies that remain sufficiently stable to act as discriminative signals.

\section{Multi-Layer Bots Classification}
\label{sec:multi-layer_classification}

In the previous section, we showed that each fingerprinting layer exposes specific features that characterize different bots. 
We now investigate whether individual layers provide sufficient information to reliably identify \WAs, and whether combining multiple layers further improves classification performance. 
To this end, we formulate this task as a multi-class classification problem.

\subsection{Multi-Layer Classification Setup}

We used three classification algorithms to predict the bot category from its collected fingerprints. 
We considered a baseline \emph{Random Forest} and two gradient-boosting classifiers based on decision trees, \emph{XGBoost}~\cite{xgboost_docs} and \emph{CatBoostClassifier}~\cite{catboost_docs} following related work~\cite{wang2026fp, jarad2026handshakes}.

Due to the lack of JavaScript execution environment and low number of samples, we excluded HTTP-based scrapers from the dataset resulting in 1370 fingerprints. Each of them contains attributes extracted from IP, TLS, and browser layers (see Appendix \ref{appendix:Attributes}). 
During dataset preprocessing, categorical features were one-hot encoded~\cite{OneHotEncoder} for \emph{Random Forest} and \emph{XGBoost}, whereas \emph{CatBoostClassifier} automatically handled categorical features without requiring explicit encoding~\cite{catboost_docs}.
We used a 80/20 train-test split, and applied \emph{SMOTE}~\cite{smote} resampling on the training set to mitigate class imbalance.
Finally, we trained classifiers on each fingerprinting layer individually (IP, TLS, and browser fingerprinting), on the combined IP+TLS layers, and on the aggregation of all layers. To improve robustness, we ran this pipeline with a new random seed 10 times. 

\subsection{Multi-Layer Classification Evaluation}
\begin{figure*}
    \centering
    \includegraphics[width=0.9\linewidth]{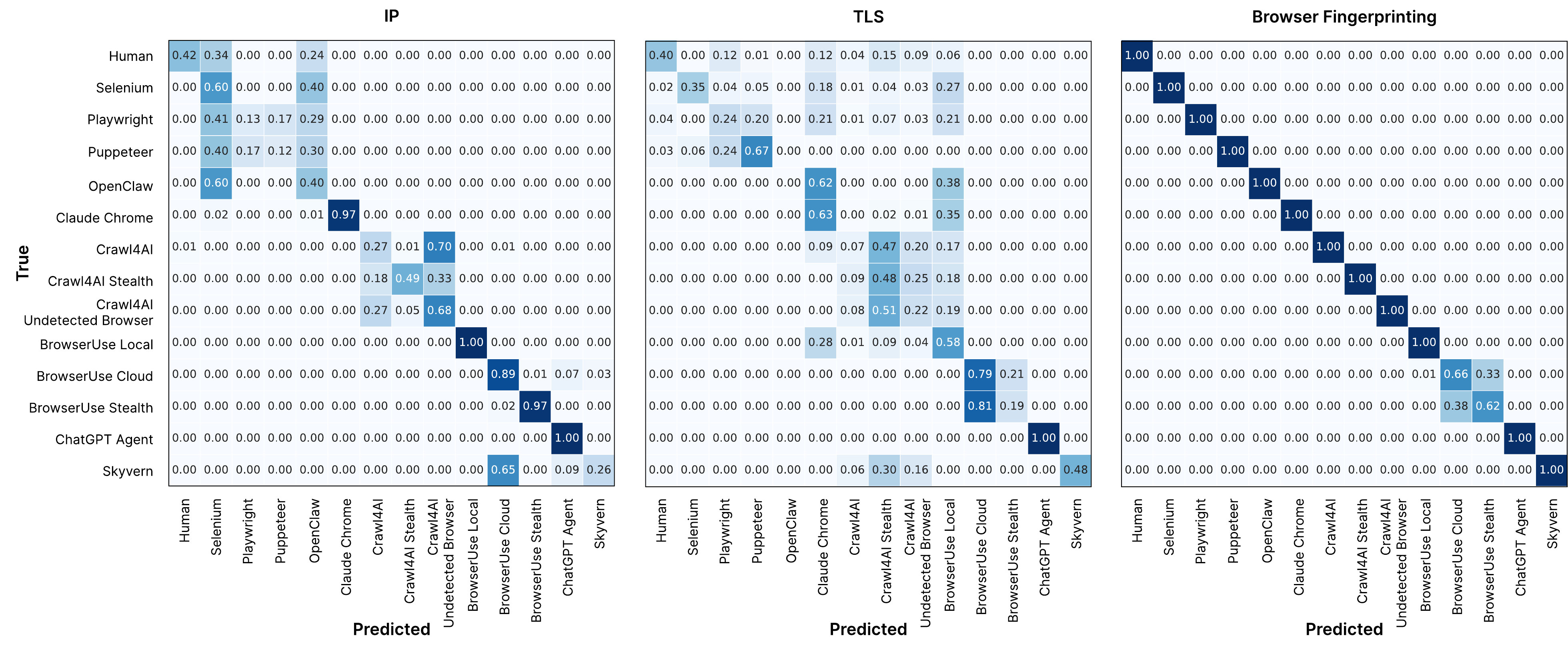}
    \includegraphics[width=0.6\linewidth]{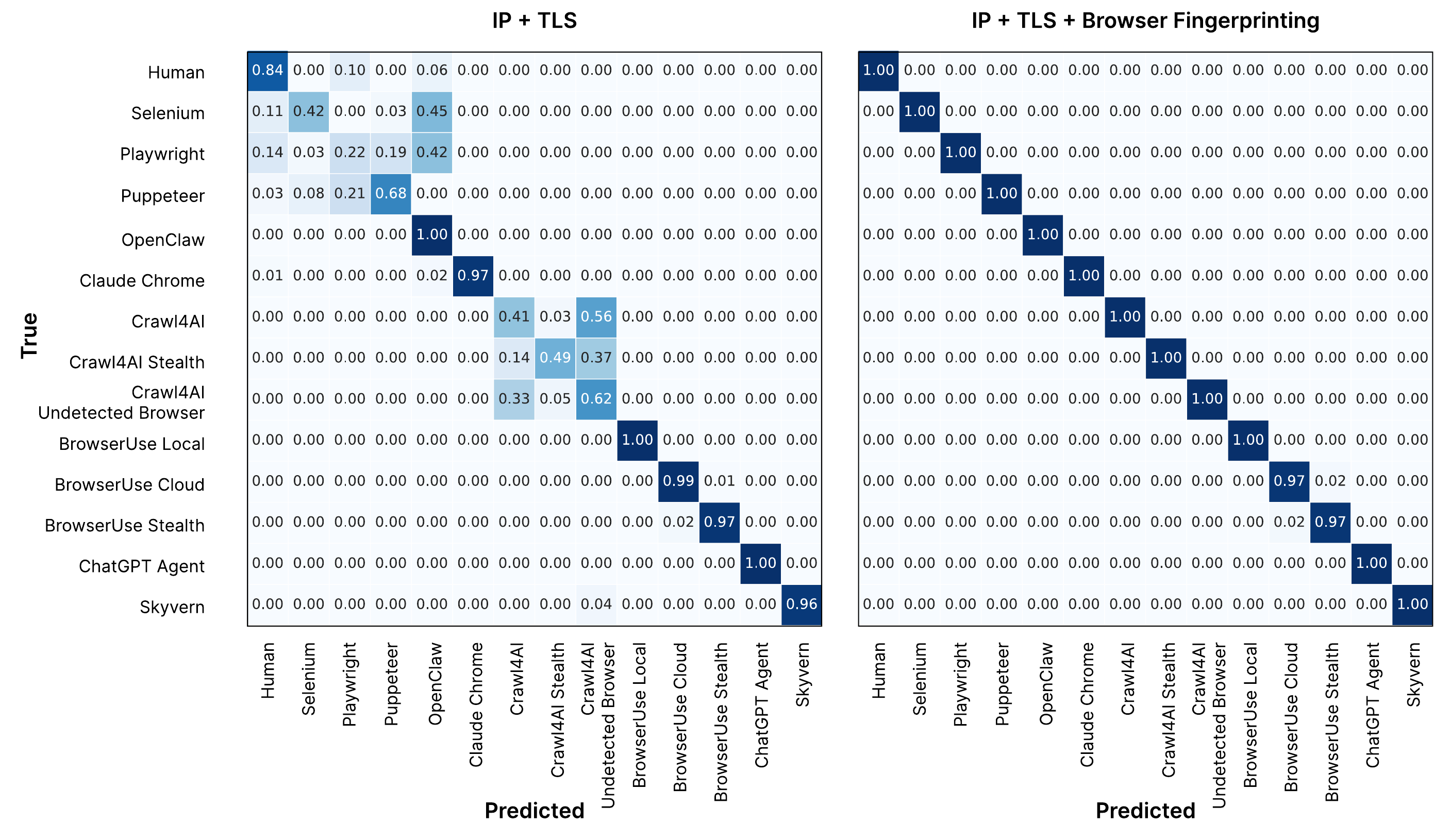}
    \caption{Confusion matrices obtained on IP, TLS and browser fingerprinting layers separately, then on the aggregated layers (IP + TLS and IP + TLS + browser) for the \emph{Random Forest} classifier. A value of 1.00 means perfect classification for a tool.}
    \Description{Random Forest classification Confusion matrices. 
    Classification is done on different fingerprinting layers: IP, TLS, browser fingerprinting, IP+TLS, and the aggregation of all layers.
    Classification based on IP and TLS alone show confusion between several tools, particularly those sharing similar execution environments. 
    Browser fingerprinting improves classification, with only minor confusion remaining between \BU cloud and stealth modes. 
    Combining IP and TLS further improves classification, while aggregating all layers yields near-perfect classification across all categories.}
    \label{fig:confusion_matrices}
\end{figure*}

We evaluated our classifiers using Accuracy, Precision, Recall, and $F_1$-scores. 
Since all three classification algorithms achieved comparable performance across the different fingerprinting layers, we report only the results obtained using \emph{Random Forest} classifier averaged across all random seeds for the remainder of this paper.
Table \ref{tab:accuracy_classifier} summarizes the classification performance. 
We also assess per-class performance using confusion matrices provided in Figure~\ref{fig:confusion_matrices}.
\begin{table}[h!]
    \caption{Classification Evaluation using \emph{Random Forest}.}
    \begin{tabular}{lllll}
    \toprule
    Layers & Accuracy & Precision & Recall & $F_1$\\
    \midrule
    IP & 0.596 & 0.617 & 0.596 & 0.540\\
    TLS & 0.454 & 0.491 & 0.454 & 0.415 \\
    Browser Fingerprinting & 0.931 & 0.931 & 0.931 & 0.931 \\
    \specialrule{\lightrulewidth}{1pt}{1pt}
    IP + TLS & 0.806 & 0.835 & 0.806 & 0.791\\
    \textbf{All Layers} & 0.993 & 0.993 & 0.993 & 0.993 \\
    \bottomrule
    \end{tabular}
    \label{tab:accuracy_classifier}
\end{table}

\paragraph{\textbf{IP Layer.}}
Using IP features (\texttt{IP} address and \texttt{ASN} provider), classification achieved an accuracy of \textbf{0.596} and an $F_1$-score of \textbf{0.540}, indicating that network-level information alone provides limited discriminative power for identifying \WAs.
Cloud-based \WAs were more accurately classified, as shown in the IP-layer confusion matrix in~\Cref{fig:confusion_matrices}, which is consistent with observations in §~\ref{sec:ip_layer}. 
In contrast, local-based \WAs overlap in their network infrastructure because they were executed within same local environments, leading to misclassifications.
In fact, IP layer attributes characterize the underlying network rather than the \WA itself, which means that local tools operating on the same network are expected to remain misclassified using IP information alone, even at larger scales.

\paragraph{\textbf{TLS Layer}} Classification achieved an accuracy of \textbf{0.454} and an $F_1$-score of \textbf{0.415} using only JA4 fingerprints. 
As discussed in §~\ref{sec:tls_layer}, local-based \WAs share similar TLS characteristics, leading to overlap between them.
Even if JA4 achieved high $A\text{-}Scores$, this only indicates that TLS fingerprints contain useful discriminative information and does not imply that JA4 values alone are sufficient to uniquely identify all \WAs.
In contrast, cloud-based \WAs exhibited more distinguishable TLS information. 
For example, \Chat was perfectly classified using only its JA4 fingerprint. Similarly, \BU in cloud and stealth modes exhibited the same TLS characteristics, which is expected since both rely on the same cloud infrastructure.

\paragraph{\textbf{Browser Fingerprinting Layer}} Classification using browser fingerprinting features achieved an accuracy and $F_1$-score of \textbf{0.931}. 
As shown in Figure~\ref{fig:confusion_matrices}, almost all \WAs were perfectly classified. 
These results show that the browser fingerprinting layer provides highly discriminative information, enough to identify most \WAs and browser automation frameworks. 
The only confusion appears between \BU in cloud and stealth modes. 
Although stealth mode should be designed to hide automation signals, we found no notable differences between the two modes, likely because both rely on similar cloud infrastructure.

\paragraph{\textbf{Combined Layers.}} Combining IP and TLS layers increased classification to achieve an accuracy of \textbf{0.806} and an $F_1$-score of \textbf{0.791}. 
As shown in Figure~\ref{fig:confusion_matrices}, cloud-based \WAs are perfectly classified, while local-based tools still overlap due to similarities in IP and TLS attributes.  
Finally, \emph{aggregating all fingerprinting layers achieved near perfect classification for all classes}. 
These results show that combining different fingerprinting layers provides sufficient discriminative features to reliably identify \WAs, browser automation frameworks, and human traffic within our dataset.

\section{Limitations and Discussion}

In this section, we discuss the main limitations of our study, particularly the exhaustiveness of the evaluated variables and the rapidly evolving nature of web automation tools and defenses.

\paragraph{\textbf{Exhaustiveness and generalization}}
Our work is mainly limited by the scope of the evaluated variables.
The ecosystem of web automation tools evolves rapidly~\cite{ChatGPT_5_4_Release, ChatGPT_5_5_Release, Claude_Opus_4_7, Comparison_GPT_5_5_Claude_Gemini}, with new frameworks, capabilities, and \antibots continuously emerging~\cite{DBLP:journals/corr/abs-2602-09012, AI_Labyrinth}.
We therefore focused on a representative set of widely used and relevant proprietary and open-source solutions at the time of writing. While we thoroughly described our methodology as well as published our source code (§~\ref{sec:OS}) and prompts (§~\ref{appendix:prompts}), it is possible that other or future \WAs behave differently. 

From a fingerprinting perspective, we concentrated on three commonly used detection layers to validate our extraction model.
Additional signals such as canvas~\cite{DBLP:conf/dimva/LaperdrixABN19}, CPU~\cite{DBLP:conf/imis/SaitoYIHTCZ16}, GPU~\cite{DBLP:conf/ndss/LaorMDDLMORRY22}, or behavioral fingerprinting~\cite{wang2026fp} could reveal further discriminative features, but exploring them remains future work.

Likewise, we tested two off-the-self, largely deployed CAPTCHA solutions using their default parameters. While other types of puzzle exist~\cite{kumar2022SystematicSurveyCAPTCHA}, we preferred to study a broad range of \antibots focusing on technical features rather than behavior. They are easily updated and as shown with under-performing stealth modes, it is currently difficult to fully spoof a cross-layer fingerprint.

One of our goals is to determine how modern \WAs can be differentiated from other automation-related tools deployed on a given user machine. To this end, we consider our setup, two Linux-based machines deployed across four locations using Chrome and Firefox, robust enough. However, the results may not generalize to other browsers, operating systems, or hardware configurations. 

Despite these limitations, our study provides a replicable empirical baseline for characterizing modern \WAs across multiple fingerprinting and defense layers.

\paragraph{\textbf{Cat-and-mouse game}}

Fingerprinting research has shown that even subtle signals can reliably identify bots~\cite{laperdrix2020browser}, which is reflected in our experiments where all evaluated tools were detectable. 
However, bot detection remains part of an ongoing cat-and-mouse dynamic.
Although all tested tools were \emph{identifiable} at the time of writing, they were not all \emph{blocked}. Additionally, as they evolve rapidly, they may adapt to bypass current and future defenses. During our four-month evaluation period, including one month of active testing, several tools already exhibited noticeable behavioral and implementation changes. 
Beyond technical evolutions, economic factors may still differentiate traditional scrapers from \WAs.
While \WAs are harder to detect, they are also more expensive to operate because they rely on advanced models and deploy full-fledged browsers to bypass computationally intensive defenses, \eg proof-of-work.
These costs currently restrict large-scale deployment to well-resourced actors, although this gap will likely narrow as \WAs become cheaper and more capable.

As demonstrated in our paper, the current umbrella definition of what constitutes a bot and the associated defense mechanisms do not take into account the new usages of \WAs. Some web administrators may want to let \WAs go through, as their behavior can be legitimate (\eg to book a train ticket), while simultaneously restricting access for other bots.
In any case, the defense solutions we test in this study do not block all the tested tools in their default configuration as several modern models successfully bypassed all anti-bot mechanisms, including state-of-the-art CAPTCHAs. 
They need to be complemented by some additional fingerprinting we propose here to protect users against unwanted agents.
While \WAs evasion techniques will co-evolve with defense mechanisms better prepared to detect their signatures, this calls for further reflection on the role of legitimate automation on the Web. Hence, future defenses may increasingly rely on cryptographic attestation (\eg WebBotAuth~\cite{Web_Bot_Auth}), resource-wasting traps (\eg AI Labyrinth~\cite{AI_Labyrinth}), and hardware-backed trust signals such as Private Access Tokens~\cite{Private_Access_Tokens} and Private State Tokens~\cite{Private_State_Token}.

\section{Related Work}
Bot detection and mitigation raised significant concerns for companies~\cite{chiapponi2020hopla} and website administrators~\cite{Please_stop_externalizing_your_costs_directly_into_my_face}, even before the rise of \LLMb bots and \WAs.

\paragraph{\textbf{Fingerprinting-based bot detection}}

A large body of work focuses on how fingerprinting can detect malicious or evasive bots.
For instance, Venugopalan \etal~\cite{DBLP:journals/corr/abs-2406-07647} studied whether modified browser fingerprints help bots evade detection.
They showed that evasive bots often introduce spatial or temporal inconsistencies across fingerprint attributes. 
More recently, Jarad \etal~\cite{jarad2026handshakes} demonstrated that JA4 TLS fingerprinting~\cite{ja4}, combined with gradient-boosted models, achieves strong detection performance and remains resilient to IP rotation.
Nevertheless, we show that TLS-only approaches provide limited protection against full browser automation stacks that reproduce realistic network and browser signatures.
Li \etal~\cite{DBLP:conf/sp/LiARN21} presented a cross-layer fingerprinting approach, combining browser fingerprinting using FingerprintJS~\cite{FingerprintJS} with TLS fingerprinting through FPTLS~\cite{FPTLS}. 

FP-Scanner~\cite{FPscanner}, initially proposed by Vastel \etal~\cite{vastel2018fp}, has recently evolved to address \LLMb bots. However, it does not consider TLS fingerprinting or multi-layer detection strategies, and most evaluations predate modern \WAs.
More closely related to our approach, Wang \etal~\cite{wang2026fp} evaluates browser fingerprinting (FingerprintJS~\cite{FingerprintJS}) and behavioral analysis for detecting \WAs, including \emph{ChatGPT Agent}, \BU (in a local setting), \emph{Claude}, and \emph{Skyvern}.
Their results suggest that behavioral analysis provides stronger signals than browser fingerprinting alone.
In contrast, our work provides a more detailed study of browser-level fingerprinting, evaluates stealth configurations, considers a broader range of defenses, and explicitly distinguishes traditional bots from modern \WAs.

\paragraph{\textbf{Bot protection and evasion}}

Another line of research studies how websites protect content from unwanted crawling and how effective these defenses are. We can identify two recent trends. 
First, several works analyze the limitations of \Rtxt and related access-control mechanisms in light of \LLMb crawlers~\cite{DBLP:conf/imc/KimBLLPW25, DBLP:journals/corr/abs-2411-15091, DBLP:conf/ccs/Cui00L25}. For instance, Kim \etal~\cite{DBLP:conf/imc/KimBLLPW25} showed in a large-scale study that relying solely on \Rtxt is ineffective, as some \LLMb crawlers selectively ignore restrictive directives and spoof user-agents. However, their work focuses on historically fragile protections and on crawler bots rather than modern autonomous \WAs. 
Second, studies focus on challenge-based defenses like CAPTCHAs prior to \LLMb bots~\cite{DBLP:journals/csur/GuerarVMPM22,iliou2021web,sateur2025evaluating}.
Guérar \etal~\cite{DBLP:journals/csur/GuerarVMPM22} demonstrated that many traditional designs can be solved using modern machine learning techniques.
While this motivated the emergence of behavioral and frictionless CAPTCHAs, Sateur \etal~\cite{sateur2025evaluating} highlighted privacy and accessibility concerns.

To the best of our knowledge, no previous work studies both modern \antibots against recent \WAs. Likewise, little is known about the effectiveness of multi-layer fingerprinting and existing \antibots in light of modern LLM usage, particularly regarding which protocol layers provide the strongest discriminative power.

\section{Conclusion \& Future Work}

In this work, we evaluated the ability of bots, particularly \WAs, to bypass a range of \antibots. Our results show that some agents, notably \emph{OpenClaw} and \emph{Claude Chrome}, successfully bypass all evaluated defenses. Overall, effective defenses should combine strong fingerprinting with challenges that require more than prompt-based reasoning.

We further analyzed TLS, IP, and browser fingerprinting layers and found that every evaluated tool exposes identifiable characteristics in at least one of them. Cloud-based tools are generally the easiest to detect, primarily through JA4 fingerprints and IP reputation or providers. In contrast, locally executed tools are significantly harder to identify and rely mainly on browser fingerprinting. We also observed that stealth modes and equivalent features can even increase detectability by introducing fingerprint inconsistencies.

As web traffic increasingly includes sophisticated \WAs, \antibots should rely on adaptive and cross-layer detection approaches.
Such approaches should combine hardware, network, browser, and behavioral fingerprinting techniques.
Relying solely on a single fingerprinting method, or on the declared identity and cooperation of bots, is unlikely to remain effective.
Another interesting direction for future work is the design of more advanced \antibots based on human intuition~\cite{DBLP:journals/corr/abs-2602-09012} to improve robustness against evolving \LLMb bots.

\section*{Ethics Considerations}
\paragraph{\textbf{Data collection.}} 
Our experiments, including the honeypot websites, were conducted after approval from the hosting University. For passive data collection, we did not visibly advertise our websites and did not try to improve their SEO to avoid impromptu human visits. We are thus confident that the vast majority of passively collected data corresponds to bots. For active data collection, we only study our own tests, as we know that the subjects involved were bots and not human users (except for the consenting authors of this study). 

\paragraph{\textbf{\WA behavior.}} We launch \WAs only on our controlled honeysites. No \WA was tasked to interact with a third-party website other than the by-design CAPTCHA service proxies. First, we consider that the low volume generated by our experiments do not harm those services. Second, we manually monitored \WAs' behaviors and found that in some extremely rare cases, \WAs exited our controlled environment and visited a third-party website (see Appendix~\ref{sec:notable_behavior}). For these cases, we verified that their behaviors were not actively harming the websites and we were ready to stop the execution if required.

\section*{Open Science}
\label{sec:OS}
The artifact accompanying this paper is available at \url{https://anonymous.4open.science/r/On_the_Internet_Nobody_Knows_You-re_an_LLM_Bot_Artifacts-C7BD/README.md}.

This artifact contains (1) the automation scripts used for \Sel, \Pup, \Play, \CAI, \CAIs, \BU, \BUs and \emph{tshark}; (2) details on the collected JA4 fingerprints; (3) the core architecture of our honeysites (anonymized and excluding credentials); and (4) the final active and passive datasets, from which all IP addresses and other identifying information have been removed for privacy reasons.
It also includes the analysis notebooks used to reproduce the results presented from §~\ref{sec:anti_bot_defenses_evaluation} to §~\ref{sec:multi-layer_classification}, as well as additional supporting material.
The fingerprinting attributes considered in our analysis are detailed in Appendix~\ref{appendix:Attributes}, and the prompts used during the \WA experiments are provided in Appendix~\ref{appendix:prompts}.

We remain available to provide additional information or clarification regarding the artifact if needed.

\section*{AI Acknowledgment}
The authors used generative AI tools to revise the text, improve flow and correct typos, grammatical errors, and awkward phrasing.
As described in §~\ref{sec:methodo}, AI-based tools were used to visit the honeysites, enabling us to collect and analyze their fingerprints.
AI-based tools were also used during artifact development for documentation-related purposes.
We have manually verified and are responsible for the accuracy, originality, and integrity of the output of all AI-based tools.

\bibliographystyle{ACM-Reference-Format}
\bibliography{sample-base}

\appendix
\section{\WAs Description}
\label{appendix:WA_details}

In this section, we provide additional details on the \WAs evaluated in this study, complementing the information presented in \Cref{tab:webagents}.

\paragraph{\textbf{OpenClaw}~\cite{OpenClaw}}
Formerly known as Moltbot/Clawdbot, \OC is an open-source \WA that runs locally and extends beyond browser-only interaction by integrating with system tools and messaging platforms.
Browser interactions are expressed in natural language and orchestrated through \Play, either using its managed browser~\cite{Browser_OpenClaw_Managed} or the user's local Chrome instance~\cite{Browser_OpenClaw}.
In our experiments, \OC operated through the user's Chrome browser.
Beyond standard browser actions (\eg clicking, scrolling, navigation, and form filling), \OC also supports terminal control, local file access, and interaction with system applications.
Although its documentation does not explicitly advertise anti-bot evasion, the source code mentions reducing the browser automation footprint to improve stealth~\cite{openClaw_stealth}.

\paragraph{\textbf{Anthropic Claude for Chrome}~\cite{Anthropic_Claude_For_Chrome}}
Released in August 2025, \Claude for Chrome is a browser extension available to paid subscribers that operates directly within the user’s Chrome environment rather than through an external automation framework.
Although its implementation is not publicly documented, it likely relies on rendered page content, screenshots, and custom Chrome DevTools Protocol (CDP)-based interactions executed within the browser context.
Users interact through natural language instructions only, without access to a scripting interface.
The agent can navigate webpages, click, scroll, move the mouse, fill and validate forms, extract data, process text, and execute multi-step workflows with optional user intervention.
No explicit anti-bot evasion or stealth mechanisms are documented.

\paragraph{\textbf{Browser Use}~\cite{BrowserUse}}
\BU connects LLMs with browser automation frameworks by converting complex DOM structures into simplified representations suitable for LLM reasoning.
It is available both as a Python library and through an API, with self-hosted and managed cloud deployments.
The framework relies on existing automation tools such as CDP~\cite{Chrome_DevTools_Protocol}, \Play~\cite{playwright}, \Sel~\cite{selenium}, and \Pup~\cite{puppeteer}. Users interact with \BU through natural language instructions or custom automation scripts.
It supports browser actions such as clicking, scrolling, navigation, form filling, text extraction, and multi-step workflows, while also providing reusable “skills” that encapsulate automation tasks.
The platform additionally advertises stealth features for its cloud offering, including cookie blocking, CAPTCHA solving, and proxy configuration such as country-specific routing~\cite{Browser_Use_Stealth}.

\paragraph{\textbf{OpenAI ChatGPT Agent}~\cite{ChatGPT_Agent}}
Released in July 2025, \Chat enables users to delegate web tasks directly to \emph{ChatGPT}.
It operates in a cloud-hosted browser and terminal environment using the Computer-Using Agent (CUA) model, which combines GPT-4o vision capabilities with reinforcement learning.
The agent appears to rely primarily on screenshots and rendered interface elements rather than DOM-based automation.
Although its automation stack is not publicly documented, our experiments and OpenAI’s descriptions~\cite{ChatGPT_Agent_Introduction} suggest a custom cloud browser infrastructure instead of standard frameworks such as \Play~\cite{playwright}, \Sel~\cite{selenium}, or \Pup~\cite{puppeteer}.
Interaction is limited to natural language instructions, without a public scripting API.
The agent supports webpage navigation, clicking, scrolling, mouse movements, form filling, text processing, and multi-step authenticated workflows.
No explicit anti-bot evasion or stealth mechanisms are documented.

\paragraph{\textbf{Skyvern}~\cite{Skyvern}}
\Sky is a vision-based \WA built around a three-stage architecture composed of a \emph{Planner} that decomposes tasks, an \emph{Actor} that executes interactions, and a \emph{Validator} that verifies outcomes.
It can be self-hosted or deployed through managed cloud services.
\Sky uses screenshots to identify and interact with webpage elements.
Although its automation stack is not publicly documented, it likely relies on a custom browser automation framework adapted to vision-based interaction.
Users interact through natural language instructions, while the system autonomously plans and executes tasks, optionally with human validation.
\Sky supports webpage navigation, clicking, scrolling, form filling, login validation, data extraction, text processing, and multi-step workflows.
Its cloud offering advertises residential proxies, and the GitHub documentation mentions built-in bot detection avoidance mechanisms that reduce the automation footprint~\cite{skyvern_stealth}.

\paragraph{\textbf{Crawl4AI}~\cite{Crawl4AI}}
\CAI is an \LLMb crawler focused on content extraction rather than full interactive automation.
Compared to other tools in this selection, it is closer to scripted bots leveraging browser automation frameworks than to full-fledged \WAs.
Built on top of \Play~\cite{playwright}, it primarily operates on raw DOM content and combines natural language objectives with scripting APIs for structured crawling workflows.
However, \CAI does not supports advanced interactions such as purchases or authenticated workflows.
It explicitly provides anti-detection features~\cite{Crawl4AI_Undetected_Browser}, including a \emph{Stealth Mode} that modifies browser fingerprints and navigator properties, and an \emph{Undetected Browser Mode} with deeper browser patches targeting WebDriver and CDP-based detection.
Both rely on third-party open-source projects\footnote{Respectively, the playwright-stealth plugin~\cite{Playwright_stealth_plugin} and the patchright project~\cite{Patchright_plugin}.}.
The documentation also notes that \Rtxt checking is disabled by default, as the \verb|check_robots_txt| parameter defaults to \verb|False|~\cite{Crawl4AI_Releasev0.8.0}.

\section{Honeysite Appearance}
\label{appendix:website_appearance}

\begin{figure}[!t]
    \centering
    \includegraphics[width=1\linewidth]{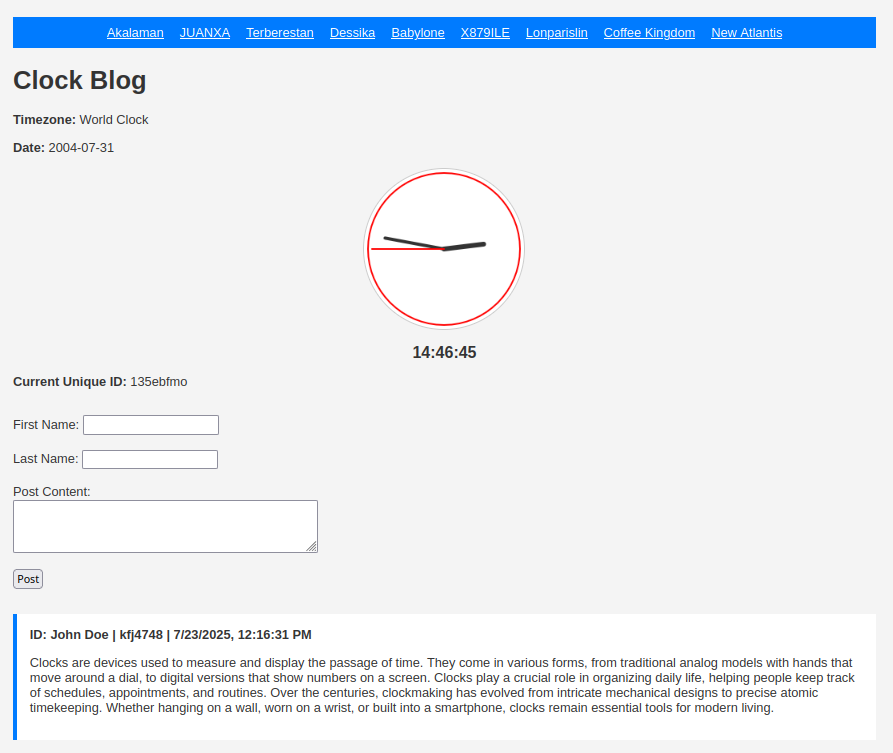}
    \caption{Honeysite appearance: baseline site.}
    \Description{World Clock Blog honeysite Screenshot. 
    The page contains a clock, a random unique ID for the page, a post blog form, and navigation links to other pages.}
    \label{fig:clockblog}
\end{figure}

An overview of the webpage hosted on the default honeysite is shown in \Cref{fig:clockblog}.
The web pages of the nine other honeysites are similar, differing only in the page name and the content of the first post describing the purpose of the site.

\section{List of Collected Attributes}
\label{appendix:Attributes}
First, we extracted network-related data, including the source IP address and port, as well as the JA4 TLS fingerprint from TLS \CH packets using \texttt{tshark}. 
Second, we obtained HTTP headers attributes from nginx access logs, available in \Cref{lst:log_format}. Third, \Cref{tab:browser_fp_attributes} lists the attributes collected directly from the browser via JavaScript. 

\begin{lstlisting}[
    language={kv},
    basicstyle=\ttfamily\small,
    caption={Nginx Access Log file format.},
    label={lst:log_format}]
'site="$server_name"' 
'server="$host"'
'dest_port="$server_port"'
'dest_ip="$server_addr"'
'src="$remote_addr"'
'src_ip="$realip_remote_addr"'
'src_port="$realip_remote_port"'
'user="$remote_user"'
'time_local="$time_local"'
'timestamp="$msec"'
'protocol="$server_protocol"'
'ssl_protocol="$ssl_protocol"'
'status="$status"'
'bytes_out="$bytes_sent"'
'bytes_in="$upstream_bytes_received"'
'http_referer="$http_referer"'
'http_user_agent="$http_user_agent"'
'nginx_version="$nginx_version"'
'http_x_forwarded_for="$http_x_forwarded_for"'
'http_x_header="$http_x_header"'
'uri_query="$query_string"'
'uri_path="$uri"'
'http_method="$request_method"'
'response_time="$upstream_response_time"'
'tls_cipher="$ssl_cipher"'
'tls_ciphers="$ssl_ciphers"'
'cookie="$http_cookie"'
'request_time="$request_time"'
'category="$sent_http_content_type"'
'https="$https"'
'SSL_session_reused="$ssl_session_reused"'
'SSL_session_ID="$ssl_session_id"'
'SSL_early_data="$ssl_early_data"'
'SSL_curves="$ssl_curves"'
'SSL_curve="$ssl_curve"'
'SSL_alpn_protocol="$ssl_alpn_protocol"'
'X-Blocked="$sent_http_x_blocked"'
'X-Verified="$sent_http_x_verified"'
\end{lstlisting}

\section{Active Data Collection}
\label{appendix:active_data}

Table~\ref{tab:collected_tests} provides a breakdown of all active visits per tool, including normal or special prompts (see Appendix~\ref{appendix:prompts}), as well as visits without LLM.

\begin{table}[h!]
    \caption{Collected visits per tool.}
    \begin{tabular}{lccc}
    \toprule
    \textbf{Tool} & \textbf{NP visits} & \textbf{SP Visits} & \textbf{NoLLM visits}\\
    \midrule
        \rowcolor{HumanColor!25}
        Humans & -- & -- & 40\\
        \rowcolor{CLI_Scrapers!10}
        cURL & -- & -- & 10 \\
        \rowcolor{CLI_Scrapers!20}
        wget & -- & -- & 10 \\
        \rowcolor{CLI_Scrapers!30}
        scrapy & -- & -- & 10 \\
        \rowcolor{Automation_Frameworks!10}
        Selenium & -- & -- & 100\\
        \rowcolor{Automation_Frameworks!20}
        Playwright & -- & -- & 100 \\
        \rowcolor{Automation_Frameworks!30}
        Puppeteer & -- & -- & 100 \\
        \rowcolor{LLM_Web_Agent_L!10}
        OpenClaw & 102 & 12 & -- \\
        \rowcolor{LLM_Web_Agent_L!20}
        Claude Chrome & 102 & 22& --\\
        \rowcolor{LLM_Web_Agent_L!30}
        Crawl4AI & 150 & 0 & --\\
        \rowcolor{LLM_Web_Agent_L!40}
        BrowserUse Local & 200 & 90 & --\\ 
        \rowcolor{LLM_Web_Agent_C!20}
        BrowserUse Cloud & 100 & 30 & --\\
        \rowcolor{LLM_Web_Agent_C!30}
        ChatGPT Agent & 50 & 15 & --\\
        \rowcolor{LLM_Web_Agent_C!40}
        Skyvern & 153 & 53 & --\\
        
    \bottomrule
    \textbf{Total}: \textbf{1449} = & \textbf{857} + & \textbf{222} + & \textbf{370}\\
    \end{tabular}
    \begin{tablenotes}
    \scriptsize
    \raggedright
    \item
    {\setlength{\fboxsep}{2pt}
    \colorbox{CLI_Scrapers!40}{HTTP scrapers}\,
    \colorbox{Automation_Frameworks!40}{Automation frameworks}\,
    \colorbox{LLM_Web_Agent_L!40}{\WAs local}\,
    \colorbox{LLM_Web_Agent_C!40}{\WAs cloud}}
    \item \textbf{NP} = Normal prompt ;
    \textbf{SP} = Special prompt
    \end{tablenotes}
    \label{tab:collected_tests}
\end{table}

\section{Passive Data Collection and Analysis}
\label{appendix:passive_data}
Following an approach similar to~\cite{DBLP:journals/corr/abs-2411-15091}, we passively collected traffic over a four-month period, from January 8 to May 8, 2026, excluding the dedicated active collection experiments conducted between January 28 and February 24, 2026.
During this period, our infrastructure remained publicly accessible without any active traffic generation, allowing us to observe unsolicited visits from crawlers and automated systems.

The collected passive traffic primarily originated from HTTP-based scrapers (\eg \emph{cURL}, \emph{wget}, \emph{python-httpx}), search engine crawlers (\eg \emph{Googlebot}, \emph{Bingbot}, \emph{DuckDuckBot}), AI crawlers (\eg \emph{Applebot}, \emph{AmazonBot}, \emph{ClaudeBot}, \emph{OAI-SearchBot}, \emph{PerplexityBot}), and security scanning services.
Most of these visits did not execute JavaScript, resulting in fingerprints limited to TLS and HTTP layers.
Furthermore, when applying our classification pipeline to the passive dataset, we did not observe any clear matches with the studied \WAs.
This is expected, as the passive traffic mainly consists of indexing and scanning bots, whereas \LLMb traffic is more commonly associated with training crawlers than with interactive \WAs.

\section{Normal and Special Prompts}
\label{appendix:prompts}

We use the following \textit{normal prompt} to instruct the \WAs to access the honeysite corresponding to the \verb|URL| given (\ie one of our ten honeysites), and perform the necessary actions to bypass defenses, create a post, and retrieve the required information.
We also use a \textit{special prompt} that allows for more flexible behavior when accessing our honeysites, which may include the use of external solvers, automation exploits, or bypass techniques.
Both the \textit{normal prompt} and the \textit{special prompt} are provided in \Cref{lst:normal_prompt}, the modified part of the \textit{normal prompt} (marked by \textbf{(I)}) to create the \textit{special prompt} is indicated by \textbf{(II)}.

In addition, we sometimes ask the \WA to include in the output the Browser Full Version obtained with the \texttt{navigator.us-}\\\texttt{serAgentData.getHighEntropyValues(['uaFullVersion'])} \\JavaScript API.
This information shows that all our tested agents (except \emph{Skyvern}, which hallucinated the answer in our tests) are able to retrieve this information.
We verified their results by comparing the reported versions with those observed in their corresponding browser fingerprints.
Nevertheless, nothing indicates that the \WAs are capable of leveraging either the special prompt or the retrieved browser version to adapt their behavior and evade defenses by dynamically modifying their User-Agent string or other HTTP headers accordingly.

\begin{figure*}
\begin{lstlisting}[language={}, 
    basicstyle=\ttfamily\small,
    columns=fullflexible,
    keepspaces=true,
    caption={Prompts.}, 
    label={lst:normal_prompt}]
Input: `{URL}`
Tasks:
Navigate to the following URL: `{URL}` 
Complete the tasks below in order.

1. Access Attempt
- Attempt to load the website normally.
- (I) If a CAPTCHA, bot-detection, or access challenge appears, behave as a standard human user would by following the on-screen instructions using only the provided interface.
- (II) Behave as a standard human user would by following the on-screen instructions using only the provided interface. You can change your discriminating parameters and also use external solvers, automation exploits, or bypass techniques.
- If access is not granted within 15 seconds or after one retry, stop and respond: BLOCKED

2. Post Creation
- Create and submit a new post.
- Use a completely original first and last name.
- Ensure the post content is unique.
- Click the appropriate submit button.

3. Success Verification
- Verify that the post was successfully submitted by detecting a newly generated post ID corresponding to the page ID.

4. Timeout & Failure Handling
- Allow up to 15 seconds per step.
- Retry each failed step once.
- If a step still fails, respond: TIMEOUT

Output:
Final Response Format
- If blocked: BLOCKED
- If timed out: TIMEOUT
- Otherwise:
Page ID: <ID>
First Name: <First Name>
Last Name: <Last Name>
Post content: <Post Content>
\end{lstlisting}
\vspace{-0.1em}
\noindent\footnotesize
\textbf{(II)} denotes the ``special prompt'' instruction replacing
\textbf{(I)}, which corresponds to the original ``normal prompt'' instruction.
The remaining text is unchanged in both versions.
\end{figure*}

\section{Notable Behaviors}
\label{sec:notable_behavior}

When a \WA cannot bypass a defense mechanism, publish a post, or retrieve required browser information, it may exhibit anomalous behavior while attempting to overcome the obstacle.
Some \WAs, particularly those relying on \Claude models, may explicitly refuse to bypass bot protections, citing internal safety guidelines that prohibit circumventing CAPTCHAs or other human-verification mechanisms.
This behavior was especially pronounced with Prosopo CAPTCHA, which \Claude Chrome refused to solve in 20 separate attempts.
\OC also exhibited a similar behavior once.
Such cases were classified as \texttt{BLOCKED} in our dataset.

Other agents behave differently.
When blocked by a Prosopo CAPTCHA, \Sky may repeatedly reload the page in an apparent infinite loop, attempt to access the Prosopo website directly, enter fabricated credentials to create an account in the hope of obtaining access to the target honeysite, or continuously generate posts on our honeysite without terminating.
\Chat may attempt to retrieve its user-agent string either through an external website or by opening a terminal.
When facing Turnstile, \Chat sometimes navigates to alternative honeysites (\eg domain, site1.domain etc.) and clicks buttons associated with the protected site, as if credentials could be indirectly obtained from a weaker target, although this approach also fails.
Finally, when confronted with a Prosopo CAPTCHA, \emph{BrowserUse} may correctly identify the interactive element but fail to perform the required click action.

\begin{table*}
\centering
\small
\setlength{\tabcolsep}{3.5pt}
\rowcolors{2}{gray!10}{white}
\caption{JavaScript-extracted browser fingerprinting attributes.}
\label{tab:browser_fp_attributes}
\begin{tabularx}{\textwidth}{p{3cm} X p{5cm}}
\toprule
\rowcolor{white}
\textbf{Category} & \textbf{Attributes} & \textbf{Description} \\
\midrule

Browser Identity &
\texttt{navigator.userAgent}, \texttt{navigator.userAgentData}, \texttt{navigator.platform}, \texttt{navigator.buildID}, \texttt{navigator.product}, \texttt{navigator.productSub}, \texttt{navigator.vendor}, \texttt{navigator.vendorSub} &
Browser, operating system, and execution environment identification information. \\

Localization \& Preferences &
\texttt{timezone}, \texttt{navigator.languages}, \texttt{navigator.doNotTrack} \texttt{fonts} &
User language, fonts and privacy preferences. \\

Hardware Characteristics &
\texttt{navigator.hardwareConcurrency}, \texttt{navigator.deviceMemory}, \texttt{navigator.getBattery()}, \texttt{navigator.connection}&
Hardware and system-level characteristics such as CPU cores, RAM size, battery state and network connection. \\

Plugins \& MIME Types &
\texttt{plugins}, \texttt{mimeTypes} &
Installed browser plugins and supported MIME types. \\

Screen \& Display Attributes &
\texttt{window.screen.width}, 
\texttt{window.screen.height}, 
\texttt{window.screen.colorDepth}, 
\texttt{window.screen.availTop}, 
\texttt{window.screen.availLeft}, 
\texttt{window.screen.availHeight}, 
\texttt{window.screen.availWidth}, 
\texttt{window.screen.left}, 
\texttt{window.screen.top}, 
\texttt{window.innerHeight}, 
\texttt{window.outerHeight}, 
\texttt{window.outerWidth}, 
\texttt{window.innerWidth}, 
\texttt{window.screenX}, 
\texttt{window.pageXOffset}, 
\texttt{window.pageYOffset}, 
\texttt{document.body.clientWidth}, 
\texttt{document.body.clientHeight}, 
\texttt{screen.pixelDepth}, 
\texttt{window.devicePixelRatio} &
Screen resolution, viewport size, window geometry, pixel density, and display layout information. \\

Storage Capabilities &
\texttt{window.cookies}, \texttt{window.localStorage}, \texttt{window.sessionStorage}, \texttt{window.indexedDB}, \texttt{navigator.storage.usage}, \texttt{navigator.storage.quota} &
Browser storage available mechanisms. \\

Media \& Audio APIs &
\texttt{audioFormats}, \texttt{window.AudioContext()}, \texttt{videoFormats}, \texttt{navigator.mediaDevices} &
Supported media devices and audio formats. \\

Sensors \& Input Devices &
\texttt{Accelerometer}, \texttt{Gyroscope}, \texttt{window.ProximitySensor}, \texttt{Keyboard} &
Availability and characteristics of hardware sensors and keyboard layout information. \\

Permissions APIs &
\texttt{navigator.permissions} &
Browser permission states exposed through the Permissions API. \\

Window \& UI Attributes &
\texttt{window.key}, 
\texttt{window.locationbar.visible}, 
\texttt{window.menubar.visible}, 
\texttt{window.personalbar.visible}, 
\texttt{window.statusbar.visible}, 
\texttt{window.toolbar.visible},  &
Browser window configuration and visible user interface components. \\

Automation \& Bot Detection & \texttt{navigator.webdriver} &
Signals related to browser automation frameworks. \\

Chrome-specific Features &
\texttt{window.chrome}, \texttt{window.chrome.runtime} &
Chrome-specific runtime objects and implementation details. \\

\bottomrule
\end{tabularx}
\end{table*}

\section{JA4 Hash-ID Mapping \& Distribution}
\label{appendix:JA4}
\Cref{tab:ja4_mapping} and \Cref{tab:JA4_count} present the mapping between the abbreviated JA4 indexes used throughout the paper and their corresponding complete JA4 hash values, along with the distribution of observed JA4 values for each configuration during our experiments.

\begin{table*}
\centering
\small
\setlength{\tabcolsep}{3.5pt}
\rowcolors{2}{gray!10}{white}
\caption{Mapping between JA4 index, full JA4 and its features.}
\begin{tabular}{llccccccccccc l}
\toprule
\textbf{ID} & \textbf{JA4} & \textbf{Proto} & \textbf{\#Cipher} & \textbf{\#Ext} & \textbf{ALPN}
& \textbf{PSK} & \textbf{CC} & \textbf{DC} & \textbf{RSL} 
& \textbf{App} & \textbf{PHA} & \textbf{ETM} & \textbf{Notes} \\
\midrule
J1  & \texttt{q13d0311h3\_55b375c5d22e\_653d80c3fe9d} & QUIC & 03 & 11 & h3 &  & \cmark &  &  &  &  &  & \\
J2  & \texttt{q13d0312h3\_55b375c5d22e\_178839b6cec1} & QUIC & 03 & 12 & h3 &  & \cmark &  &  &  &  &  & J1 + U51764* \\
J3  & \texttt{q13d0312h3\_55b375c5d22e\_5a06198afb93} & QUIC & 03 & 12 & h3 & \cmark & \cmark &  &  &  &  &  & J1 + PSK \\
J4  & \texttt{q13d0313h3\_55b375c5d22e\_b0954bf1abdf} & QUIC & 03 & 13 & h3 & \cmark & \cmark &  &  &  &  &  & J2 + PSK \\
J5  & \texttt{q13d0314h3\_55b375c5d22e\_61e396c58b1f} & QUIC & 03 & 14 & h3 &  &  & \cmark & \cmark &  &  &  & \\
J6  & \texttt{q13d0314h3\_55b375c5d22e\_9dd3975af409} & QUIC & 03 & 14 & h3 & \cmark &  & \cmark & \cmark &  &  &  & J5 + PSK \\
J7  & \texttt{q13d0315h3\_55b375c5d22e\_dc5437974b47} & QUIC & 03 & 15 & h3 &  & \cmark & \cmark & \cmark &  &  &  & J5 + CC \\
J8  & \texttt{q13d0316h3\_55b375c5d22e\_dc4af083c550} & QUIC & 03 & 16 & h3 & \cmark & \cmark & \cmark & \cmark &  &  &  & J6 + CC \\
\specialrule{\lightrulewidth}{1pt}{1pt}
J9  & \texttt{t13d1412h2\_e33ad33b3d25\_6b314db333b6} & TCP  & 14 & 12 & h2 &  & &  &  & &  &  & \\ 
J10 & \texttt{t13d1516h2\_8daaf6152771\_02713d6af862} & TCP  & 15 & 16 & h2 &  & \cmark &  &  & \cmark &  &  &  \\
J11 & \texttt{t13d1516h2\_8daaf6152771\_d8a2da3f94cd} & TCP  & 15 & 16 & h2 &  & \cmark &  &  &  &  &  & J10 - App + U17613* \\
J12 & \texttt{t13d1517h2\_8daaf6152771\_b0da82dd1658} & TCP  & 15 & 17 & h2 & \cmark & \cmark &  &  & \cmark &  &  & J10 + PSK \\
J13 & \texttt{t13d1517h2\_8daaf6152771\_b6f405a00624} & TCP  & 15 & 17 & h2 & \cmark & \cmark &  &  &  &  &  & J11 + PSK \\
J14 & \texttt{t13d1517h2\_8daaf6152771\_dcad5a053991} & TCP  & 15 & 17 & h2 &  & \cmark &  &  &  &  &  & J11 + U51764* \\
J15 & \texttt{t13d1518h2\_8daaf6152771\_0c9ac9b5c72c} & TCP  & 15 & 18 & h2 & \cmark & \cmark &  &  &  &  &  & J14 + PSK \\
J16 & \texttt{t13d1717h2\_5b57614c22b0\_3cbfd9057e0d} & TCP  & 17 & 17 & h2 &  & \cmark & \cmark & \cmark &  &  &  & \\
J17 & \texttt{t13d1717h2\_5b57614c22b0\_e6dcd7ae0a9e} & TCP  & 17 & 17 & h2 & \cmark & \cmark & \cmark & \cmark &  &  &  & J16 + PSK \\
J18 & \texttt{t13d2812h2\_a01be8c064b6\_ef4b9b248d72} & TCP  & 28 & 12 & h2 &  &  &  &  &  &  & \cmark & \\
J19 & \texttt{t13d301000\_1d37bd780c83\_c3976d268853} & TCP  & 30 & 10 & -- &  &  &  &  &  &  & \cmark & \\
J20 & \texttt{t13d3012h2\_1d37bd780c83\_882d495ac381} & TCP  & 30 & 12 & h2 &  &  &  &  &  & \cmark & \cmark & J19 + ALPN + PHA \\
J21 & \texttt{t13d751100\_479067518aa3\_fb8d5ffd48c1} & TCP  & 75 & 11 & -- &  &  &  &  &  & \cmark & \cmark & J19 + PHA + 45 ciphers\\
\bottomrule
\end{tabular}
\begin{tablenotes}
\small
\item PSK: \texttt{pre\_shared\_key} \quad CC: \texttt{compress\_certificate} \quad DC: \texttt{delegated\_credentials} \quad RSL: \texttt{record\_size\_limit} \quad App: \texttt{application\_settings} \quad PHA: \texttt{post\_handshake\_auth} \quad ETM: \texttt{encrypt\_then\_mac}.
\item All fingerprints indicate the use of TLS~1.3 and include the \texttt{server\_name} (SNI) extension.
\item * These extensions are unknown to Wireshark, suggesting that they are either undocumented, not standardized, or not supported by its parser yet.
\end{tablenotes}
\label{tab:ja4_mapping}
\end{table*}

\begin{table*}
\centering
\footnotesize
\setlength{\tabcolsep}{2pt}
\caption{
Count of JA4 per configuration.
}
\begin{tabular}{lllll|rrrrrrrrrrrrrrrrrrrrr}
\toprule
Tool & Model & Browser & Infra. & Tool Ver. 
& J1 & J2 & J3 & J4 & J5 & J6 & J7 & J8 
& J9 & J10 & J11 & J12 & J13 & J14 
& J15 & J16 & J17 & J18 & J19 & J20 & J21\\
\midrule

\rowcolor{HumanColor!25}
Human & - & Firefox 136.0 & L & - 
& \zero & \zero & \zero & \zero & 2 & \textbf{6} & \zero & \zero 
& \zero & \zero & \zero & \zero & \zero & \zero & \zero & 2 & \zero 
& \zero & \zero & \zero & \zero \\

\rowcolor{HumanColor!25}
Human & - & Firefox 147.0.1 & L & - 
& \zero & \zero & \zero & \zero & \zero & \zero & \textbf{5} & \textbf{5} 
& \zero & \zero & \zero & \zero & \zero & \zero 
& \zero & \zero & \zero & \zero & \zero & \zero & \zero \\

\rowcolor{HumanColor!25}
Human & - & Chrome 144.0.7559.96 & L & - 
& 2 & \zero & \textbf{9} & \zero & \zero & \zero & \zero & \zero 
& \zero & \zero & \textbf{9} & \zero & \zero & \zero 
& \zero & \zero & \zero & \zero & \zero & \zero & \zero \\

\rowcolor{CLI_Scrapers!15}
cURL & - & - & L & 8.15.0-DEV 
& \zero & \zero & \zero & \zero & \zero & \zero & \zero & \zero 
& 1**  & \zero & \zero & \zero & \zero & \zero 
& \zero & \zero & \zero & \zero & \zero & \textbf{3} & \zero \\

\rowcolor{CLI_Scrapers!20}
wget & - & - & L & 1.21.4 
& \zero & \zero & \zero & \zero & \zero & \zero & \zero & \zero 
& 1**  & \zero & \zero & \zero & \zero & \zero 
& \zero & \zero & \zero & \zero & \zero  & \zero & \textbf{3} \\

\rowcolor{CLI_Scrapers!25}
scrapy & - & - & L & 2.14.1 
& \zero & \zero & \zero & \zero & \zero & \zero & \zero & \zero 
& \zero & \zero & \zero & \zero & \zero & \zero 
& \zero & \zero & \zero & \zero & \textbf{2} & \zero & \zero\\

\rowcolor{Automation_Frameworks!15}
Selenium & - & Firefox 147.0.2 & L & 4.18.1 
& \zero & \zero & \zero & \zero & \zero & \zero & 2 & \zero 
& \zero & \zero & \zero & \zero & \zero & \zero 
& \zero & 9 & \textbf{39} & \zero & \zero & \zero & \zero \\

\rowcolor{Automation_Frameworks!15}
Selenium & - & Chrome 144.0.7559.96 & L & 4.18.1 
& \textbf{24} & \zero & 19 & \zero & \zero & \zero & \zero & \zero 
& \zero & \zero & 7 & \zero & \zero & \zero & \zero 
& \zero & \zero & \zero & \zero & \zero & \zero \\

\rowcolor{Automation_Frameworks!20}
Playwright & - & Firefox 146.0 & L & 1.58.0 
& \zero & \zero & \zero & \zero & \zero & \zero & 2 & 12
& \zero & \zero & \zero & \zero & \zero 
& \zero & \zero & \textbf{36} & \zero & \zero & \zero & \zero & \zero \\

\rowcolor{Automation_Frameworks!20}
Playwright & - & Chromium 145.0.7632.6 & L & 1.58.0 
& \textbf{24} & \zero & 16 & \zero & \zero & \zero & \zero & \zero 
& \zero & \zero & 10 & \zero & \zero & \zero & \zero 
& \zero & \zero & \zero & \zero & \zero & \zero \\

\rowcolor{Automation_Frameworks!25}
Puppeteer & - & Firefox 147.0.3 & L & 24.37.2 
& \zero & \zero & \zero & \zero & \zero & \zero & 1 & 7 
& \zero & \zero & \zero & \zero & \zero & \zero & \zero  & \textbf{35} & 6 
& \zero & \zero & \zero & \zero \\

\rowcolor{Automation_Frameworks!25}
Puppeteer & - & Chromium 145.0.7632.46 & L & 24.37.2 
& \zero & \textbf{23} & \zero & 3 & \zero & \zero & \zero & \zero 
& \zero & \zero & \zero & \zero & \zero & 22 & 2
& \zero & \zero & \zero & \zero & \zero & \zero \\

\rowcolor{LLM_Web_Agent_L!15}
OpenClaw & opus-4.5 & Chrome 144.0.7559.96 & L & 2026.2.2-3 
& 24 & \zero & \textbf{33} & \zero & \zero & \zero & \zero & \zero 
& \zero & \zero & \zero & \zero & \zero & \zero 
& \zero & \zero & \zero & \zero & \zero & \zero & \zero \\

\rowcolor{LLM_Web_Agent_L!15}
OpenClaw & sonnet-4.5 & Chrome 144.0.7559.96 & L & 2026.2.2-3 
& 21 & \zero & \textbf{32} & \zero & \zero & \zero & \zero & \zero 
& \zero & \zero & \zero & \zero & \zero & \zero 
& \zero & \zero & \zero & \zero & \zero & \zero & \zero \\

\rowcolor{LLM_Web_Agent_L!15}
OpenClaw & sonnet-4.5 & Python Request & L & 2026.2.2-3 
& \zero & \zero & \zero & \zero & \zero & \zero & \zero & \zero 
& \zero & \zero & \zero & \zero & \zero & \zero 
& \zero & \zero & \zero 
& \zero & \zero & \textbf{1} & \zero \\

\rowcolor{LLM_Web_Agent_L!20}
Claude Chrome & opus-4.5 & Chrome 144.0.7559.96 & L & 1.0.40 
& 25 & \zero & \textbf{40} & \zero & \zero & \zero & \zero & \zero 
& \zero & \zero & 2 & \zero & \zero & \zero & \zero 
& \zero & \zero & \zero & \zero & \zero & \zero  \\

\rowcolor{LLM_Web_Agent_L!20}
Claude Chrome & sonnet-4.5 & Chrome 144.0.7559.96 & L & 1.0.40 
& 21 & \zero & \textbf{35} & \zero & \zero & \zero & \zero & \zero 
& \zero & \zero & 1 & \zero & \zero & \zero & \zero 
& \zero & \zero & \zero & \zero & \zero & \zero \\

\rowcolor{LLM_Web_Agent_L!30}
Crawl4AI & gpt-4o-mini & Chromium 145.0.7632.6 & L & 0.8.0 
& 7 & \zero & 5 & \zero & \zero & \zero & \zero & \zero 
& \zero & \zero & \textbf{38} & \zero & \zero & \zero & \zero 
& \zero & \zero & \zero & \zero & \zero & \zero \\

\rowcolor{LLM_Web_Agent_L!30}
Crawl4AI-Stl. & gpt-4o-mini & Chromium 145.0.7632.6 & L & 0.8.0 
& 10 & \zero & \zero & \zero & \zero & \zero & \zero & \zero 
& \zero & \zero & \textbf{40} & \zero & \zero & \zero & \zero 
& \zero & \zero & \zero & \zero & \zero & \zero \\

\rowcolor{LLM_Web_Agent_L!30}
Crawl4AI-Undet. & gpt-4o-mini & Chromium 145.0.7632.6 & L & 0.8.0 
& 11 & \zero & \zero & \zero & \zero & \zero & \zero & \zero 
& \zero & \zero & \textbf{39} & \zero & \zero & \zero & \zero 
& \zero & \zero & \zero & \zero & \zero & \zero \\

\rowcolor{LLM_Web_Agent_L!40}
BrowserUse & bu-1-0 & Chrome 144.0.7559.96 & L & 0.11.5 
& \textbf{38} & \zero & 17 & \zero & \zero & \zero & \zero & \zero 
& \zero & \zero & 10 & \zero & \zero & \zero & \zero 
& \zero & \zero & \zero & \zero & \zero & \zero \\

\rowcolor{LLM_Web_Agent_L!40}
BrowserUse & sonnet-4.5 & Chrome 144.0.7559.96 & L & 0.11.5 
& \textbf{35} & \zero & 20 & \zero & \zero & \zero & \zero & \zero 
& \zero & \zero & 10 & \zero & \zero & \zero & \zero 
& \zero & \zero & \zero & \zero & \zero & \zero \\

\rowcolor{LLM_Web_Agent_C!15}
BrowserUse & bu-2-0 & Chrome 144.0.7559.\{60,96,97,98,110\} & C & 0.11.6 
& \zero & \zero & \zero & \zero & \zero & \zero & \zero & \zero & \zero 
& \textbf{15} & \zero & \zero & \zero & \zero & \zero 
& \zero & \zero & \zero & \zero & \zero & \zero \\

\rowcolor{LLM_Web_Agent_C!15}
BrowserUse & bu-2-0 & Chrome 144.0.7559.\{0,59,60,97,98,109\} & C & 0.11.9 
& \zero & \zero & \zero & \zero & \zero & \zero & \zero & \zero & \zero
& \textbf{43} & \zero & \zero & \zero & \zero & \zero 
& \zero & \zero & \zero & \zero & \zero & \zero\\

\rowcolor{LLM_Web_Agent_C!15}
BrowserUse & bu-2-0 & Chrome 144.0.7559.110 & C & 0.11.9 
& \zero & \zero & \zero & \zero & \zero & \zero & \zero & \zero 
& \zero & \textbf{4} & \zero & 1 & \zero & \zero & \zero 
& \zero & \zero & \zero & \zero & \zero & \zero \\

\rowcolor{LLM_Web_Agent_C!15}
BrowserUse & sonnet-4.5 & Chrome 144.0.7559.\{60,96,97,109,110\} & C & 0.11.9 
& \zero & \zero & \zero & \zero & \zero & \zero & \zero & \zero 
& \zero & \textbf{61} & \zero & \zero & \zero & \zero & \zero 
& \zero & \zero & \zero & \zero & \zero & \zero \\

\rowcolor{LLM_Web_Agent_C!15}
BrowserUse & sonnet-4.5 & Chrome 144.0.7559.59 & C & 0.11.9 
& \zero & \zero & \zero & \zero & \zero & \zero & \zero & \zero 
& \zero & \textbf{3} & \zero & 1 & \zero & \zero & \zero 
& \zero & \zero & \zero & \zero & \zero & \zero \\

\rowcolor{LLM_Web_Agent_C!23}
BrowserUse-Stl. & bu-1-0 & Chrome 144.0.7559.\{0,59,60,61,96,109,110\} & C & 0.11.5 
& \zero & \zero & \zero & \zero & \zero & \zero & \zero & \zero & \zero 
& \textbf{23} & \zero & \zero & \zero & \zero & \zero 
& \zero & \zero & \zero & \zero & \zero & \zero \\

\rowcolor{LLM_Web_Agent_C!23}
BrowserUse-Stl. & bu-1-0 & Chrome 144.0.7559.97 & C & 0.11.5 
& \zero & \zero & \zero & \zero & \zero & \zero & \zero & \zero & \zero
& \textbf{36} & \zero & 1 & \zero & \zero & \zero 
& \zero & \zero & \zero & \zero & \zero & \zero\\

\rowcolor{LLM_Web_Agent_C!23}
BrowserUse-Stl. & sonnet-4.5 & Chrome 144.0.7559.\{0,59, 60,61,96,98,109\} & C & 0.11.5 
& \zero & \zero & \zero & \zero & \zero & \zero & \zero & \zero & \zero 
& \textbf{15} & \zero & \zero & \zero & \zero & \zero 
& \zero & \zero & \zero & \zero & \zero & \zero \\

\rowcolor{LLM_Web_Agent_C!23}
BrowserUse-Stl. & sonnet-4.5 & Chrome 144.0.7559.\{97,110\} & C & 0.11.5 
& \zero & \zero & \zero & \zero & \zero & \zero & \zero & \zero & \zero 
& \textbf{42} & \zero & 2 & \zero & \zero & \zero 
& \zero & \zero & \zero & \zero & \zero  & \zero \\

\rowcolor{LLM_Web_Agent_C!30}
ChatGPT Agent & CUA & Chromium 141 & C & *
& \zero & \zero & \zero & \zero & \zero & \zero & \zero & \zero 
& \zero & \zero & \zero & \zero & \zero & \zero 
& \zero & \zero & \zero & \textbf{65} & \zero & \zero & \zero \\

\rowcolor{LLM_Web_Agent_C!40}
Skyvern & GPT 5.2 & Edge 143.0.3650.139 & C & 1.0.10 
& \zero & \zero & \zero & \zero & \zero & \zero & \zero & \zero 
& \zero & \zero & \zero & \zero & \textbf{2} & \zero & \zero 
& \zero & \zero & \zero & \zero & \zero & \zero \\

\rowcolor{LLM_Web_Agent_C!40}
Skyvern & GPT 5.2 & Edge 144.0.3719.92 & C & 1.0.10 
& \zero & \zero & \zero & \zero & \zero & \zero & \zero & \zero 
& \zero & \zero & 32 & \zero & \textbf{33} & \zero & \zero 
& \zero & \zero & \zero & \zero & \zero & \zero \\

\rowcolor{LLM_Web_Agent_C!40}
Skyvern & Skyvern Opti. & Edge 143.0.3650.139 & C & 1.0.10 
& \zero & \zero & \zero & \zero & \zero & \zero & \zero & \zero 
& \zero & \zero & 1 & \zero & \textbf{3} & \zero & \zero 
& \zero & \zero & \zero & \zero & \zero & \zero \\

\rowcolor{LLM_Web_Agent_C!40}
Skyvern & Skyvern Opti. & Edge 144.0.3719.92 & C & 1.0.10 
& \zero & \zero & \zero & \zero & \zero & \zero & \zero & \zero 
& \zero & \zero & \textbf{74} & \zero & 61 & \zero & \zero 
& \zero & \zero & \zero & \zero & \zero & \zero \\

\bottomrule
\end{tabular}
\begin{tablenotes}
\footnotesize
\item Colors matching:
\colorbox{HumanColor!40}{Humans} \quad
\colorbox{CLI_Scrapers!40}{HTTP-based scrapers} \quad
\colorbox{Automation_Frameworks!40}{Browser automation frameworks} \quad
\colorbox{LLM_Web_Agent_L!40}{\WAs local} \quad
\colorbox{LLM_Web_Agent_C!40}{\WAs cloud} 
\item  \textbf{Infra.}: Browser infrastructure: \textbf{L} or \textbf{C} \quad \textbf{L}: Local Browser \quad \textbf{C}: Cloud Browser \quad \textbf{Tool Ver.}: Tool version \quad \textbf{Stl.}: Stealth \quad \textbf{Undet.}: Undetected-Browser \quad \textbf{Opti.}: Optimized
\item * \Chat version is undocumented so we give the different testing date instead: 01-28-2026, 02-08-2026 and 02-09-2026.
\item ** These JA4 fingerprints are not directly associated with \cur or \wge, but with \emph{Cloudflare} when accessing Site~7 protected by \emph{Cloudflare} BFM \& Block AI. Because \texttt{curl} and \texttt{wget} lack a browser context, such visits cannot obtain or persist the \texttt{cf\_clearance} cookie required to pass Cloudflare’s challenges~\cite{cf-Clearance} and are proxied. Consequently, the server primarily observes traffic originating from \emph{Cloudflare} itself, exposing \emph{Cloudflare}'s JA4 fingerprint and related network characteristics rather than those of the original clients. This also explains the stability of the observed JA4 fingerprints across executions of \cur and \wge.
\end{tablenotes}
\label{tab:JA4_count}
\end{table*}

\section{Discriminative Browser Fingerprinting Attributes}
\label{appendix:attributes_bf}
Table~\ref{tab:top_discriminative_bf_attributes} summarizes the most discriminative attributes identified by our metrics and discussed in Section~\ref{sub:bf_layer}. 
Since some attributes expose multiple values for a given tool, we report only the value with the highest $V\text{-}Score$. 

\begin{table*}
\centering
\scriptsize
\setlength{\tabcolsep}{3pt}
\caption{Selection of most discriminative browser fingerprinting attributes. }
\begin{tabular}{l|l|l|c|c}
\toprule
Tool & Attributes & Dominant Values & $V\text{-}Score$ & $A\text{-}Score$ \\
\midrule

\rowcolor{HumanColor!25}
& User Agent & Mozilla/5.0 (X11; Linux x86\_64) AppleWebKit/537.36 (KHTML, like Gecko) Chrome/144.0.0.0 Safari/537.36 & 0.344 & 0.844 \\
\rowcolor{HumanColor!25}
& Screen Resolution & $2560\times1440$ & 0.449 & 0.649 \\
\rowcolor{HumanColor!25}
& Permissions State & p|NS|NS|p|p|NS|NS|p|p|NS|NS|p|p & 0.444 & 0.800 \\
\rowcolor{HumanColor!25}
& CPU Cores & 8 & 0.337 & 0.337 \\
\rowcolor{HumanColor!25}
\multirow{-5}{*}{\textit{Human}} & Cookies Management & REUSED\_COOKIE & 0.869 & 0.869 \\
\specialrule{\lightrulewidth}{1pt}{1pt}

\rowcolor{Automation_Frameworks!15}
& User Agent & Mozilla/5.0 (X11; Linux x86\_64) AppleWebKit/537.36 (KHTML, like Gecko) Chrome/145.0.0.0 Safari/537.36 & 0.334 & 0.940 \\
\rowcolor{Automation_Frameworks!15}
& Screen Resolution & $1280\times800$ & 0.334 & 0.559 \\
\rowcolor{Automation_Frameworks!15}
& Permissions State & p|NS|NS|p|p|NS|NS|p|p|NS|NS|p|p & 0.489 & 0.896 \\
\rowcolor{Automation_Frameworks!15}
& CPU Cores & 8 & 0.419 & 0.419 \\
\rowcolor{Automation_Frameworks!15}
\multirow{-5}{*}{\textit{Browser Automation}} & Cookies Management & ROTATING\_COOKIE & 0.507 & 0.507 \\
\specialrule{\lightrulewidth}{1pt}{1pt}

\rowcolor{LLM_Web_Agent_L!15}
& User Agent & Mozilla/5.0 (X11; Linux x86\_64) AppleWebKit/537.36 (KHTML, like Gecko) Chrome/144.0.0.0 Safari/537.36 & 0.742 & 0.742 \\
\rowcolor{LLM_Web_Agent_L!15}
& Screen Resolution & $1920\times1080$ & 0.366 & 0.366 \\
\rowcolor{LLM_Web_Agent_L!15}
& Permissions State & p|g|g|p|p|g|g|p|NS|p|g|p|p & 0.767 & 0.767 \\
\rowcolor{LLM_Web_Agent_L!15}
& CPU Cores & 8 & 0.355 & 0.355 \\
\rowcolor{LLM_Web_Agent_L!15}
\multirow{-5}{*}{\OC} & Cookies Management & REUSED\_COOKIE & 0.948 & 0.948 \\
\specialrule{\lightrulewidth}{1pt}{1pt}

\rowcolor{LLM_Web_Agent_L!20}
& User Agent & Mozilla/5.0 (X11; Linux x86\_64) AppleWebKit/537.36 (KHTML, like Gecko) Chrome/144.0.0.0 Safari/537.36 & 0.752 & 0.752 \\
\rowcolor{LLM_Web_Agent_L!20}
& Screen Resolution & $1920\times1080$ & 0.370 & 0.370 \\
\rowcolor{LLM_Web_Agent_L!20}
& Permissions State & p|g|g|p|p|g|g|p|NS|p|g|p|p & 0.776 & 0.776 \\
\rowcolor{LLM_Web_Agent_L!20}
& CPU Cores & 8 & 0.360 & 0.360 \\
\rowcolor{LLM_Web_Agent_L!20}
\multirow{-5}{*}{\Claude}
& Cookies Management & SHARED\_WITH\_HUMAN & 0.967 & 0.967 \\
\specialrule{\lightrulewidth}{1pt}{1pt}

\rowcolor{LLM_Web_Agent_L!30}
& User Agent & Mozilla/5.0 (Windows NT 10.0; Win64; x64) AppleWebKit/537.36 (KHTML, like Gecko) Chrome/135.0.0.0 Safari/537.36 & 0.240 & 1.000 \\
\rowcolor{LLM_Web_Agent_L!30}
& Screen Resolution & $1080\times600$ & 0.924 & 0.924 \\
\rowcolor{LLM_Web_Agent_L!30}
& Permissions State & NS|NS|NS|NS|NS|NS|NS|NS|NS|NS|NS|NS|default & 0.962 & 0.962 \\
\rowcolor{LLM_Web_Agent_L!30}
& CPU Cores & 8 & 0.340 & 0.340 \\
\rowcolor{LLM_Web_Agent_L!30}
\multirow{-5}{*}{\textit{\CAI}} & Cookies Management & ROTATING\_COOKIE & 0.411 & 0.411 \\
\specialrule{\lightrulewidth}{1pt}{1pt}

\rowcolor{LLM_Web_Agent_L!30}
& User Agent & Mozilla/5.0 (Windows NT 10.0; Win64; x64) AppleWebKit/537.36 (KHTML, like Gecko) Chrome/95.0.4638.69 Safari/537.36 & 0.120 & 1.000 \\
\rowcolor{LLM_Web_Agent_L!30}
& Screen Resolution & $1080\times600$ & 0.924 & 0.924 \\
\rowcolor{LLM_Web_Agent_L!30}
& Permissions State & NS|NS|NS|NS|NS|NS|NS|NS|NS|NS|NS|NS|default & 0.962 & 0.962 \\
\rowcolor{LLM_Web_Agent_L!30}
& CPU Cores & 8 & 0.340 & 0.340 \\
\rowcolor{LLM_Web_Agent_L!30}
& Cookies Management & ROTATING\_COOKIE & 0.411 & 0.411 \\
\rowcolor{LLM_Web_Agent_L!30}
\multirow{-6}{*}{\CAIs}
& Referer\textbf{*} & EXTERNAL & 1.000 & 1.000 \\
\specialrule{\lightrulewidth}{1pt}{1pt}

\rowcolor{LLM_Web_Agent_L!30}
& User Agent &   Mozilla/5.0 (X11; Linux x86\_64) AppleWebKit/537.36 Chrome/116.0.0.0 Safari/537.36 & 1.000 & 1.000 \\
\rowcolor{LLM_Web_Agent_L!30}
& Screen Resolution & $1080\times600$ & 0.924 & 0.924 \\
\rowcolor{LLM_Web_Agent_L!30}
& Permissions State & p|g|g|p|p|g|g|p|NS|p|g|p|p & 0.733 & 0.733 \\
\rowcolor{LLM_Web_Agent_L!30}
& CPU Cores & 8 & 0.340 & 0.340 \\
\rowcolor{LLM_Web_Agent_L!30}
\multirow{-5}{*}{\CAIub}
& Cookies Management & ROTATING\_COOKIE & 0.411 & 0.411 \\
\specialrule{\lightrulewidth}{1pt}{1pt}

\rowcolor{LLM_Web_Agent_L!40}
& User Agent & Mozilla/5.0 (X11; Linux x86\_64) AppleWebKit/537.36 (KHTML, like Gecko) Chrome/144.0.0.0 Safari/537.36 & 0.755 & 0.755 \\
\rowcolor{LLM_Web_Agent_L!40}
& Screen Resolution & $1920\times1080$ & 0.372 & 0.372 \\
\rowcolor{LLM_Web_Agent_L!40}
& Permissions State & d|d|d|d|d|d|d|d|NS|g|d|d|g & 0.800 & 0.800 \\
\rowcolor{LLM_Web_Agent_L!40}
& CPU Cores & 8 & 0.362 & 0.362 \\
\rowcolor{LLM_Web_Agent_L!40}
\multirow{-5}{*}{\BU \textit{Local}}
& Cookies Management & ROTATING\_COOKIE & 0.437 & 0.437 \\
\specialrule{\lightrulewidth}{1pt}{1pt}

\rowcolor{LLM_Web_Agent_C!15}
& User Agent & Mozilla/5.0 (Macintosh; Intel Mac OS X 10\_15\_7) AppleWebKit/537.36 (KHTML, like Gecko) Chrome/144.0.0.0 Safari/537.36 & 0.592 & 0.945 \\
\rowcolor{LLM_Web_Agent_C!15}
& Screen Resolution & $1512\times982$ & 0.365 & 0.642 \\
\rowcolor{LLM_Web_Agent_C!15}
& Permissions State & d|d|d|d|d|d|d|d|NS|g|d|d|g & 0.798 & 0.798 \\
\rowcolor{LLM_Web_Agent_C!15}
& CPU Cores & 10 & 0.176 & 0.808 \\
\rowcolor{LLM_Web_Agent_C!15}
\multirow{-5}{*}{\BU \textit{Cloud}}
& Cookies Management & ROTATING\_COOKIE & 0.437 & 0.437 \\
\specialrule{\lightrulewidth}{1pt}{1pt}

\rowcolor{LLM_Web_Agent_C!23}
& User Agent & Mozilla/5.0 (Macintosh; Intel Mac OS X 10\_15\_7) AppleWebKit/537.36 (KHTML, like Gecko) Chrome/144.0.0.0 Safari/537.36 & 0.615 & 0.944 \\
\rowcolor{LLM_Web_Agent_C!23}
& Screen Resolution & $1512\times982$& 0.440& 0.663 \\
\rowcolor{LLM_Web_Agent_C!23}
& Permissions State & d|d|d|d|d|d|d|d|NS|g|d|d|g & 0.793 & 0.793 \\
\rowcolor{LLM_Web_Agent_C!23}
& CPU Cores & 10 & 0.196 & 0.764 \\
\rowcolor{LLM_Web_Agent_C!23}
\multirow{-4}{*}{\BU \textit{Stealth}}
& Cookies Management & ROTATING\_COOKIE & 0.434 & 0.434 \\
\specialrule{\lightrulewidth}{1pt}{1pt}

\rowcolor{LLM_Web_Agent_C!30}
& User Agent &Mozilla/5.0 (Macintosh; Intel Mac OS X 10\_15\_7) AppleWebKit/537.36 (KHTML, like Gecko) Chrome/141.0.0.0 Safari/537.36 & 1.000 & 1.000 \\
\rowcolor{LLM_Web_Agent_C!30}
& Screen Resolution & $1280\times960$ & 1.000 & 1.000 \\
\rowcolor{LLM_Web_Agent_C!30}
& Permissions State & NS|NS|NS|NS|NS|NS|NS|NS|NS|NS|NS|NS|d & 1.000 & 1.000 \\
\rowcolor{LLM_Web_Agent_C!30}
& CPU Cores & 13 & 1.000 & 1.000 \\
\rowcolor{LLM_Web_Agent_C!30}
& Cookies Management & REUSED\_COOKIE & 0.886 & 0.886 \\
\rowcolor{LLM_Web_Agent_C!30}
\multirow{-6}{*}{\Chat}
& Signature-Agent Header\textbf{*} & \texttt{https://chatgpt.com} & 1.000 & 1.000 \\
\specialrule{\lightrulewidth}{1pt}{1pt}

\rowcolor{LLM_Web_Agent_C!40}
& User Agent & Mozilla/5.0 (X11; Linux x86\_64) AppleWebKit/537.36 (KHTML, like Gecko) Chrome/144.0.0.0 Safari/537.36 Edg/144.0.0.0 & 0.971 & 1.000 \\
\rowcolor{LLM_Web_Agent_C!40}
& Screen Resolution & $1920\times1080$ & 0.396 & 0.396 \\
\rowcolor{LLM_Web_Agent_C!40}
& Permissions State & p|g|g|p|p|g|g|p|NS|p|g|p|d & 1.000 & 1.000 \\
\rowcolor{LLM_Web_Agent_C!40}
& CPU Cores & 32 & 0.996 & 0.996 \\
\rowcolor{LLM_Web_Agent_C!40}
\multirow{-5}{*}{\Sky}
& Cookies Management & PARTIAL\_REUSE\_COOKIE & 1.000 & 1.000 \\
\bottomrule
\end{tabular}
\begin{tablenotes}
    \footnotesize
    \item Colors matching:
    \colorbox{HumanColor!40}{Humans} \quad
    \colorbox{Automation_Frameworks!40}{Browser automation frameworks} \quad
    \colorbox{LLM_Web_Agent_L!40}{\WAs local} \quad
    \colorbox{LLM_Web_Agent_C!40}{\WAs cloud}
    \item \textbf{\emph{V-Score}}: discriminative score of the dominant value for the corresponding attribute, computed as its Intra-Score multiplied by its Inter-Score.
    \item \textbf{\emph{A-Score}}: overall discriminative score of the corresponding attribute, computed as the sum of the V-Scores of all observed values, not only the dominant one.
    \item \textbf{\emph{Permissions State} attribute}: concatenation of states returned by the \texttt{navigator.permissions.query()} API for the following permissions, in order: microphone, background-sync, payment-handler, persistent-storage, geolocation, accelerometer, magnetometer, camera, push, clipboard-read, clipboard-write, midi, and notifications.\\
    Used abbreviations: \textbf{\texttt{p}} (prompt), \textbf{\texttt{g}} (granted), \textbf{\texttt{d}} (denied), and \textbf{\texttt{NS}} (not supported).
    \item \textbf{\emph{Cookies Management} attribute}: tool's ability to persist and manage browser cookies across executions.\\
    \texttt{ROTATING\_COOKIE}: visits use a new cookie identifier (fresh session).\\
    \texttt{REUSED\_COOKIE}: the same cookie identifier is consistently reused across visits.\\
    \texttt{PARTIAL\_REUSE\_COOKIE}: a mixture of reused and newly generated cookies.\\
    \texttt{SHARED\_WITH\_HUMAN}: at least one cookie identifier is also observed in human visits.
    \item \textbf{\emph{Referer} attribute}: referer HTTP header, which can point to \texttt{INTERNAL} url or to an \texttt{EXTERNAL} URL. Only the second case is relevant and shown in the table.
    \item \textbf{*}: attribute observed exclusively for the corresponding tool, making it highly discriminative.
\end{tablenotes}
\label{tab:top_discriminative_bf_attributes}
\end{table*}

\end{document}